\documentclass{elsart}

\usepackage{cite,graphicx,amsmath,amsfonts,amssymb,mathrsfs,bbm}

\begin{document}

\begin{frontmatter}

\begin{flushright}
{\small TUM-HEP-498/03}
\end{flushright}

\title{Deconstruction and Bilarge Neutrino Mixing}

\author[labela]{Gerhart Seidl\thanksref{label1}}
\thanks[label1]{E-mail: gseidl@ph.tum.de}

\address[labela]{Institut f{\"u}r Theoretische Physik, Physik-Department,
Technische Universit{\"a}t M{\"u}nchen, James-Franck-Stra{\ss}e, 85748
Garching bei M{\"u}nchen, Germany}

\date{\today}
         
\begin{abstract}
We present a model for lepton masses and mixing angles using the deconstruction
setup, based on a non-Abelian flavor symmetry which is a split extension of the
Klein group $Z_2\times Z_2$. The symmetries enforce an approximately
maximal atmospheric mixing angle $\theta_{23}$, a nearly vanishing
reactor mixing angle $\theta_{13}$, and the hierarchy of charged lepton masses.
The charged lepton mass spectrum emerges from the Froggatt-Nielsen
mechanism interpreted in terms of deconstructed extra dimensions compactified
on $\mathcal{S}^1$. A normal neutrino mass hierarchy arises from the coupling to
right-handed neutrinos propagating in latticized $\mathcal{S}^1/Z_2$ orbifold
extra dimensions. Here, the solar mass squared difference $\Delta m_\odot^2$ is
suppressed by the size of the dynamically generated bulk manifold. Without
tuning of parameters, the model yields the solar mixing angle
$\theta_{12}={\rm arctan}\:1/\sqrt{2}$. Thus, the obtained 
neutrino masses and mixing angles are all in agreement with the
Mikheyev--Smirnov--Wolfenstein large mixing angle solution of the solar
neutrino problem.
\end{abstract}

\end{frontmatter}

\newpage
\section{Introduction}
The standard model of elementary particle physics (SM), minimally extended by
massive Majorana neutrinos, contains 22 fermion mass and mixing parameters
(6 quark masses, 6 lepton masses, 3 CKM mixing angles \cite{cab63,kob73},
3 MNS mixing angles \cite{maki62}, 2 Dirac $\mathcal{CP}$ violation phases,
and 2 Majorana phases). In four-dimensional
(4D) gauge theories the fermion masses and mixing angles result from Yukawa
couplings between fermions and scalars. However, even in 4D grand
unified theories (GUTs) the Yukawa couplings are described by many free
parameters
and one therefore often assumes that the structure of
the Yukawa coupling matrices is dictated by a sequentially broken
flavor symmetry. Indeed, since the 4D Yukawa couplings become calculable
in higher dimensional gauge theories \cite{wett85}, the effective fermion mass
matrices emerging after dimensional reduction should actually be
highly predictable from symmetries or the topology of internal space
\cite{strom85}.

At first sight, the hierarchical structure of CKM angles and charged
fermion masses suggests an underlying non-Abelian flavor symmetry group which
essentially acts only on the first and second generations
\cite{barb97}.
However, in these theories it is difficult\footnote{For a possible
counter-example, see the recent model in Ref. \cite{moha002}.} to obtain exact
predictions compatible with recent atmospheric \cite{super98,super0012}
and solar \cite{super001,sno001}
neutrino experimental results which prefer the Mikheyev-Smirnov-Wolfenstein
(MSW) \cite{mik85} large mixing angle (LMA) solution of the solar
neutrino problem \cite{bah0021}. In fact, the
KamLAND \cite{kam002} reactor neutrino experiment has recently confirmed the
MSW LMA solution at a significant level \cite{barg0022,fo002,mal002,bah0022}. In
the ``standard'' parameterization, the MSW LMA solution tells us that the
leptons obey a {\it bilarge} mixing in which the solar mixing angle $\theta_{12}$
is large, but not close to maximal, the atmospheric mixing angle $\theta_{23}$
is close to maximal, and the reactor mixing angle $\theta_{13}$ is small.

By assuming only Abelian $U(1)$ \cite{el97} or $Z_n$
\cite{gro98} symmetries one finds that the
atmospheric mixing angle may be large but cannot be enforced to be close to
maximal. Thus, a natural close to maximal $\nu_\mu$-$\nu_\tau$-mixing can be
interpreted as a strong hint for some non-Abelian flavor symmetry acting on the
2nd and 3rd generations \cite{moha99,wett99,king01}.
Neutrino mass models which give large or maximal solar
and atmospheric mixing angles by assigning the 2nd and 3rd generations quantum
numbers of the symmetric groups $S_2$ \cite{grim001} or $S_3$ \cite{moha000}
have, in general, difficulties to address the hierarchy of the charged lepton
masses. This problem can be resolved in a supersymmetric model for
degenerate neutrino masses by imposing the group $A_4$, the symmetry
group of the tetrahedron \cite{babu002}. In addition, this model predicts
exactly 
$\theta_{23}=\frac{\pi}{4}$ from the $A_4$ symmetry and gives with some tuning of
parameters the solar mixing angle $\theta_{12}$ of the MSW LMA solution.
However, in unified field theories a normal hierarchical neutrino mass spectrum
seems more plausible than an inverted or degenerate one \cite{king0021}.
A comparably simple way of accommodating the MSW LMA solution for normal
hierarchical neutrino masses is provided in scenarios of
single right-handed neutrino dominance \cite{king99}.

Although unification in more than four dimensions can serve as a motivation for
flavor symmetries, higher-dimensional gauge theories have dimensionful gauge
couplings and are usually non-renormalizable. They require a truncation
of Kaluza-Klein (KK) modes near some cut-off scale $M_f$ at which
the perturbative regime of the higher-dimensional gauge theory breaks down.
Recently, however, a new class of 4D gauge-invariant field theories for
deconstructed or latticized extra dimensions has been proposed, which generate
the physics of extra dimensions in their infrared limit \cite{ark001,hill001}.
These theories are renormalizable and can thus be viewed as viable UV
completions of some fundamental non-perturbative field theory, like string theory.
In deconstruction, a number of replicated gauge groups is connected by fermionic
or bosonic link variables. Since replicated gauge groups also frequently appear
in models using the Froggatt-Nielsen mechanism
\cite{frog79,frog94}, it is straightforward to account for the
generation of fermion mass matrices by deconstructed extra dimensional gauge
symmetries.

In two previous works \cite{ohl0021,ohl0022} we have presented versions of a
model for bilarge leptonic mixing based on a vacuum alignment mechanism for
a strictly hierarchical charged lepton mass spectrum and an inverted
neutrino mass hierarchy. As a result of the inverse hierarchical neutrino mass
spectrum, we obtained the approximate relation
$\theta_{12}=\frac{\pi}{4}-\theta_{13}$ and hence the lower bound
$37^{\circ}\lesssim\theta_{12}$ on the solar mixing angle. In this paper,
we now apply this vacuum alignment mechanism to a model for
lepton masses and mixing angles which yields more comfortably the MSW LMA
solution with normal hierarchical neutrino masses. Specifically, the lepton mass
hierarchies are generated by deconstructed extra-dimensional $U(1)$ gauge
symmetries where the lattice-link variables are themselves subject to a
discrete non-Abelian flavor symmetry.

The paper is organized as follows: in Section \ref{sec:deconstruction}, we
introduce our example field theory by briefly reviewing the periodic and the
aliphatic model for deconstructed extra dimensions. In Section
\ref{sec:generators}, we first assign the horizontal $U(1)$ charges, then we
discuss the non-Abelian discrete flavor symmetry and examine its normal
structure. Next, in Section \ref{sec:scalarpotential},
we construct the potential for the scalar link-variables of the latticized extra
dimensions from the representations of the dihedral group $\mathscr{D}_4$.
The vacuum structure which results from minimizing the scalar potential is
determined and analyzed in Section \ref{sec:vacuumalignment}.
In Section \ref{sec:chargedleptons}, we describe the generation of
the charged lepton masses via the vacuum alignment mechanism.
Then, in Section \ref{sec:neutrinos}, we study the
neutrino mass matrix, determine the neutrino mass and mixing parameters and 
match the types of latticizations of the orbifold extra dimensions
onto the presently allowed ranges for $\Delta m_\odot^2$ implied by recent
KamLAND results. Finally, we present in Section 
\ref{sec:summaryandconclusions} our summary and conclusions.
In addition, Appendix \ref{app:dihedral}
gives a brief review of the dihedral group $\mathscr{D}_4$ and in Appendix
\ref{app:minimization} the minimization of the scalar
potential is explicitly carried out.

\section{Deconstruction setups}\label{sec:deconstruction}
\subsection{The periodic model}\label{sec:periodic}
\begin{figure}
\begin{center}
\includegraphics*[bb = 235 595 340 680]{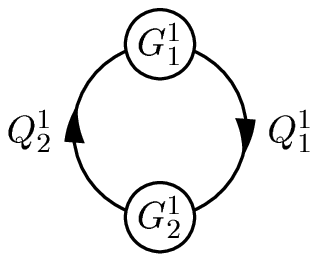}
\end{center}
\vspace*{-3mm}
\caption{\small{Moose diagram for the gauge group $G^1=\Pi^2_{i=1}G^1_i$.}}
  \label{fig:moose2}
\end{figure}

\begin{figure}
\begin{center}
\includegraphics*[bb = 230 570 345 680]{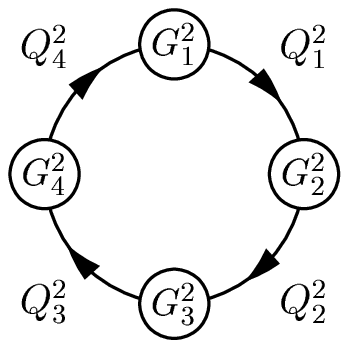}
\end{center}
\vspace*{-3mm}
\caption{\small{Moose diagram for the gauge group $G^2=\Pi_{i=1}^4 G^2_i$.}}
  \label{fig:moose1}
\end{figure}

Consider a 4D gauge-invariant field theory for deconstructed or
latticized extra dimensions \cite{ark001,hill001}. Let the field theory for
$j=1,2,\ldots$ be described by products $G^j=\Pi^{N_j}_{i=1}G^j_i$ of
$U(1)$ gauge groups $G^j_i$, where $N_j$ is the total number of
``sites'' corresponding to $G^j$.
In the ``periodic model'' \cite{ark001}, one associates with each pair of
groups $G^j_i\times G^j_{i+1}$ a link-Higgs field $Q^j_i$ with charge $(+1,-1)$
under $G^j_i\times G^j_{i+1}$, where $i$ is periodically identified with
$i+N_j$. For our example field theory we will assume one periodic model for
each of the gauge groups $G^1$ and $G^2$ where the number of sites is
$N_1=2$ and $N_2=4$.
This field theory is
conveniently represented by the
``moose'' \cite{geor86} or ``quiver'' \cite{doug96} diagrams in
Figs.~\ref{fig:moose2} and \ref{fig:moose1}.
Restricting for the present to a single product gauge group $G^j$ with $N_j=N$
sites, we can drop the index $j$ and write the Lagrangian in the periodic model
as
\begin{equation}\label{eq:L_N}
 {\mathscr{L}}=-\frac{1}{4}\sum_{i=1}^N F_{i\mu\nu}F^{i\mu\nu}
 +\sum_{i=1}^N\left(D_\mu Q_i\right)^\dagger D^\mu Q_i,
\end{equation}
where $F_{i\mu\nu}=\partial_\mu A_{i\nu}-\partial_\nu A_{i\mu}$ is the
4D field strength and the covariant derivative
is defined by
\begin{equation}\label{eq:covariantderivative}
D_\mu Q_i=(\partial_\mu-{\rm i}g_iA_{i\mu}+{\rm i}g_{i+1}A_{(i+1)\mu})
Q_i.
\end{equation}
where
$g_i$ is the dimensionless coupling constant of the gauge group $G$ at the $i$th
site.
For simplicity, it is assumed that all gauge couplings
are equal,
\begin{equation}
 g_1=g_2=\ldots =g_N\equiv g.
\end{equation}
When the $Q_i$ aquire after spontaneous symmetry breaking (SSB)
universal vacuum expectation values (VEVs) and each field becomes a non-linear
$\sigma$-model field
\begin{equation}
 Q_i\rightarrow v\:{\rm exp}({\rm i}\alpha_i/v),
\end{equation}
the full gauge symmetry is higgsed to the diagonal $U(1)$ and it is seen that the
Lagrangian in Eq.~(\ref{eq:L_N}) describes a 4+1 dimensional $U(1)$ gauge
theory on a flat background, where only the fifth dimension has been latticized.
The lattice spacing
$a$ and circumference $R$ of the fifth dimension are $a=1/(gv)$ and
$R=N/(gv)$, whereas the 5D gauge coupling $g_5$ is given by 
$1/g_5^2=1/(ag^2)$ \cite{ark001,ark002}.
(In the model with free boundary conditions \cite{hill001} generic gauge couplings and non-universal VEVs introduce an overall
non-trivial warp factor \cite{cheng001,fal002}, resulting, {\it e.g.}, in a
Randall-Sundrum model \cite{ran99}.)
Implicitly, we identify the link-Higgs fields with the continuum Wilson lines
\begin{equation}
 Q_i(x^\mu)={\rm exp}\left({\rm i}\int_{ia}^{(i+1)a}{{\rm d}x^5\:
 A_5(x^\mu,x^5)}\right),
\end{equation}
where $(x^\mu,x^5)$ are the bulk coordinates and $A_5$ is the fifth
component of the bulk gauge field associated with the gauge group $G$.
The gauge boson mass matrix takes the form
\begin{equation}
 \sum_{i=1}^N g^2v^2(A_{i\mu}-A_{(i+1)\mu})(A_{i}^\mu-A_{i+1}^\mu)
\end{equation}
and is of the type of a nearest neighbor
coupled oscillator Hamiltonian. In the case of $N=4$, for example,
the mass matrix for the gauge bosons reads \begin{equation}
 g^2v^2
 \left(
 \begin{matrix}
  2 &-1& 0 & -1\\
  -1 & 2 & -1 & 0\\
  0 & -1 & 2 & -1\\
  -1 & 0 & -1 & 2
 \end{matrix}
 \right).
\end{equation}
In general \cite{ark001}, the gauge boson mass matrix yields a mass
spectrum labeled by an integer $k$ satisfying $-N/2<k\leq N/2$,
\begin{equation}
 M_k^2=4g^2v^2{\rm sin}^2\left(\frac{\pi k}{N}\right)
 \equiv\left(\frac{2}{a}\right)^2{\rm sin}^2\left(\frac{p_5a}{2}\right),
\end{equation}
where $p_5\equiv 2\pi k/R$ is the discrete 5D momentum. For small $|k|$ the
masses are
\begin{equation}
 M_k\simeq |p_k|=\frac{2\pi |k|}{R},\quad |k|\ll N/2,
\end{equation}
which is exactly the KK spectrum\footnote{Note here the doubling of KK modes.}
of a 5D gauge theory compactified on $\mathcal{S}^1$ with circumference
$R=N/(gv)$. The zero mode corresponds
to the unbroken, diagonal $U(1)$. Hence, at energies
$E\ll gv/N$ we observe an ordinary 4D gauge theory, in the range
$gv/N<E< gv$ the physics is that of an extra dimension, and for $E\gg
gv$ an unbroken $U(1)^N$ gauge theory in four dimensions is recovered.

\subsection{The aliphatic model for fermions}\label{sec:aliphatic}
\begin{figure}
\begin{center}
\includegraphics*[bb = 155 601 420 687]{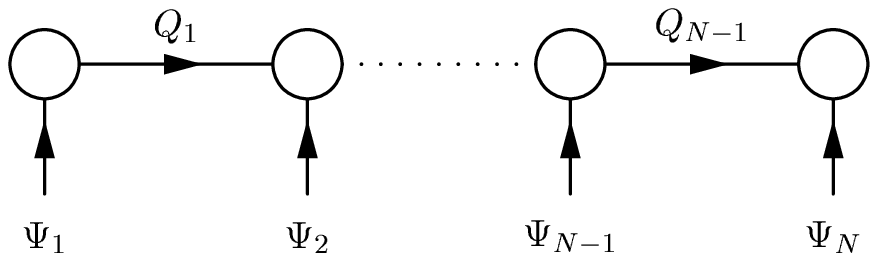}
\end{center}
\vspace{-3mm}
\caption{\small{Moose diagram for bulk vector fields and fermions in the
  aliphatic model.}}
  \label{fig:aliphatic2}
\end{figure}

The ``aliphatic model'' \cite{hill001} for some product gauge group
$G^j$ $(j=3,4,\ldots)$, as defined in Sec.~\ref{sec:periodic}, is obtained from
the periodic model by setting $Q^j_{N_j}=0$ which yields a linear system with
free boundary conditions.
For this type of latticization we will consider $N_j$ SM singlet Dirac fermions
$\Psi^j_n\:(n=1,\ldots ,N_j)$, each of which carries 
an associated $G^j_n$ charge $-1$. The SM singlet Higgs fields
$Q^j_n\:(n=1,\ldots N_j)$ which are assigned the $G^j_n\times G^j_{n+1}$ charges
$(+1,-1)$ specify the allowed couplings between the fermions. Restricting here 
to a single product gauge group $G^j$ we may in this section
drop the index $j$ for the rest of the discussion. The moose diagram for
the aliphatic model is shown in Fig.~\ref{fig:aliphatic2}.
We will denote by $\Psi_{nL}$ and $\Psi_{nR}$ the left- and right-handed
chiral components of $\Psi_n$ respectively. To engineer chiral fermion zero modes
from compactification of the 5th dimension, one can impose discretized
versions of the Neumann and
Dirichlet boundary conditions $\Psi_{NL}-\Psi_{(N-1)L}=0$ and
$\Psi_{1R}=\Psi_{NR}=0$ which explicitly break the Lorentz group
in five dimensions. Using the transverse lattice technique
\cite{bard76,bard80} the relevant mass and mixing terms of $\Psi_{nL}$ and
$\Psi_{mR}$ then read \cite{hill001,cheng001} 
\begin{eqnarray}\label{eq:L_1}
{\mathscr{L}}_1 & = & M_f
\sum_{n=2}^{N-1}\left[\overline{\Psi}_{nL}\left(\frac{Q^\dagger_{n+1}}{v}
\Psi_{(n+1)R}-\Psi_{nR}\right)
-\overline{\Psi}_{nR}\left(\Psi_{nL}-\frac{Q_n}{v}\Psi_{(n-1)L}\right)\right]
\nonumber\\
&&+M_f\overline{\Psi}_{1L}\Psi_{2R}+{\rm h.c.},
\end{eqnarray}
where $v\equiv\langle Q_n\rangle$ for $n=1,\ldots ,N-1$, {\it i.e.}, we have universal
VEVs. In fact, Eq.~(\ref{eq:L_1}) is the Wilson-Dirac action \cite{wil74} for a
transverse extra dimension which reproduces the 5D continuum theory in the limit
of vanishing lattice spacing. Clearly, this tacitly presupposes a specific
functional measure for the link variables in question which may, however, no
longer be a necessary choice when the lattice spacing is finite \cite{bard80}.
After SSB the mixed mass terms of the chiral fermions are
\begin{eqnarray}
{\mathscr{L}}_{\rm mass} & = & M_f
  \sum_{n=2}^{N-1}\left[\overline{\Psi}_{nL}(\Psi_{(n+1)R}-\Psi_{nR})
  -\overline{\Psi}_{nR}(\Psi_{nL}-\Psi_{(n-1)L})
  \right]\nonumber\\
 &&+M_f\overline{\Psi}_{1L}\Psi_{2R}+{\rm h.c.}\nonumber\\
 &=& (\overline{\Psi}_{1L},\ldots,\overline{\Psi}_{(N-1)L})^T {\mathcal{M}}_1
     (\Psi_{2R},\ldots,\Psi_{(N-1)R})+{\rm h.c.},
\end{eqnarray}
where the $(N-1)\times (N-2)$ fermion mass matrix ${\mathcal{M}}_1$ is on the form
\begin{equation}
 {\mathcal{M}}_1=M_f
 \left(
  \begin{matrix}
   1 & 0 & 0& \cdots & 0 & 0\\
   -1 & 1 & 0 & \cdots & 0 & 0\\
   \vdots & \vdots & \vdots & &\vdots & \vdots\\
   0 & 0 & 0 & \cdots & -1 & 1\\
   0 & 0 & 0 & \cdots & 0 & - 1
  \end{matrix}
  \right).
\end{equation}
Diagonalizing the $(N-2)\times (N-2)$ matrix ${\mathcal{M}}_1^\dagger{\mathcal{M}}_1$
gives for the mass eigenvalues of the right-handed fermions
\begin{eqnarray}\label{eq:M_nR}
 M_{nR}=2M_f\:{\rm sin}\left(\frac{n\pi}{2(N-1)}\right),\quad n=1,2,\ldots,N-2,
\end{eqnarray}
where the associated mass eigenstates $\tilde{\Psi}_{kR}$ are related to
$\Psi_{nL}$ by \cite{hill001}:
\begin{equation}
 \Psi_{nR}=\sqrt{\frac{N-1}{2}}\sum_{k=1}^{N-2}{\rm sin}\left(n\frac{k\pi}{N-1}
 \right)\tilde{\Psi}_{kR}.
\end{equation}
The diagonalization of the $(N-1)\times (N-1)$ matrix ${\mathcal{M}}_1{\mathcal{M}}_1^\dagger$
yields masses for the left-handed fermions, which are identical with the gauge boson masses,
\begin{equation}\label{eq:M_n,L}
 M_{nL}=2M_f\:{\rm sin}\left(\frac{n\pi}{2(N-1)}\right),\quad n=0,1,\ldots,N-2,
\end{equation}
and in terms of the associated the mass eigenstates $\tilde{\Psi}_{kL}$ we
have \cite{hill001}:
\begin{equation}
 \Psi_{nL}=\sqrt{\frac{N-1}{2}}\sum_{k=0}^{N-2}{\rm cos}\left(\frac{2n+1}{2}
 \frac{k\pi}{N-1}\right)\tilde{\Psi}_{kL}.
\end{equation}
Setting $M_f=\frac{N-1}{R}$ and taking the limit  $n\ll N$ the linear KK
spectrum of a 5D fermion in an orbifold extra dimension $\mathcal{S}^1/Z_2$ is reproduced.
Up to an overall scale factor of 2, the periodic and the aliphatic model
generate identical KK mass spectra for bulk vector fields, with the number of
KK modes doubled in the periodic case \cite{cheng0012}.

\section{Horizontal charges}\label{sec:generators}
In 5D continuum theories hierarchical Yukawa coupling matrix textures have been
successfully generated \cite{gher000}. Hierarchical Yukawa matrices have
also been obtained in deconstructed warped geometries \cite{Abe:2002rj}. In our
deconstruction setup of Sec.~\ref{sec:deconstruction} let us now consider an
extension of the SM, where the lepton masses arise from higher-dimensional
operators \cite{wilc79}, partly via the Froggatt-Nielsen mechanism.

\subsection{$U(1)$ charges}\label{sec:U(1)charges}
We denote the left-handed SM lepton doublets as
$L_\alpha =(\nu_{\alpha L}\:e_{\alpha L})^T$ and the
right-handed charged leptons as $E_\alpha$, where $\alpha =e,\mu,\tau$ is the
flavor index. For simplicity, we will assume the electroweak scalar sector to
consist only of the SM Higgs doublet $H$. In order to account for the
neutrino masses, we will additionally assume three SM
singlet scalar fields $\xi_0,\xi_1,$ and $\xi_2,$ as well as three heavy SM singlet Dirac
neutrinos $F_e,F_\mu,$ and $F_\tau$. Since these Dirac neutrinos are supposed
to have masses of the order of the GUT scale $\simeq 10^{15}\:{\rm GeV}$,
they will account for the smallness of the neutrino masses via the seesaw mechanism
\cite{yana79,gell79}. Furthermore, the standard Froggatt-Nielsen mechanism
is implied in the charged lepton sector by the presence of heavy fundamental
charged fermion messengers. The electron, muon, and tau masses will be denoted
by $m_e,m_\mu,$ and $m_\tau$, respectively.

In models of inverted neutrino mass hierarchy, a small reactor mixing angle
$\theta_{13}$ can be understood in terms of a softly broken
$\tilde{L}_e-\tilde{L}_\mu-\tilde{L}_\tau$ lepton number \cite{pet82}.
Analogously, we assume that our example field theory contains
a product $G_L\times G_R$ of two extra $U(1)$ gauge symmetries $G_L$ and $G_R$
which distinguish the 1st generation from the 2nd and 3rd generations, but act 
also on
$\xi_0,\xi_1,\xi_2,$ and the scalar link fields of the product gauge groups
$G^1=\Pi^{2}_{i=1}G^1_i$ and $G^2=\Pi^4_{i=1}G^2_i$.
The $G_L\times G_R$ charge assignment is shown in table \ref{tab:GLGRcharges}.
\begin{table}
\begin{center}
\begin{tabular}{lc}
\hline
&\vspace{-0.7cm}\\
\vspace{0.5mm}&$G_L\times G_R$\\
\hline
\vspace{-0.5cm}&\\
$L_e$&$(+2,-2)$\\
$E_e,F_e,\xi_0$& $(-2,+2)$\\
$L_\mu,L_\tau,Q^1_1,Q^2_1,\xi_1,\xi_2$&$(+1,0)$\\
$F_\mu,F_\tau$&$(-1,0)$\\
$E_\mu,E_\tau$&$(0,+1)$\\
$Q^1_2,Q^2_4$&$(0,-1)$\\
\vspace{-3.5mm}&\\
\hline
\end{tabular}
\end{center}
\caption{Assignment of the horizontal $U(1)^2$ charges to the different fields.
The fields not shown here transform trivially under $G_L\times G_R$.}
\label{tab:GLGRcharges}
\end{table}
Note that the $G_L\times G_R$ symmetry is anomalous. However, anomalous
$U(1)$ charges are expected to arise in string theory and must be canceled by the
Green-Schwarz mechanism \cite{green84}.

\begin{figure}
\begin{center}
\includegraphics*[bb = 112 557 475 650]{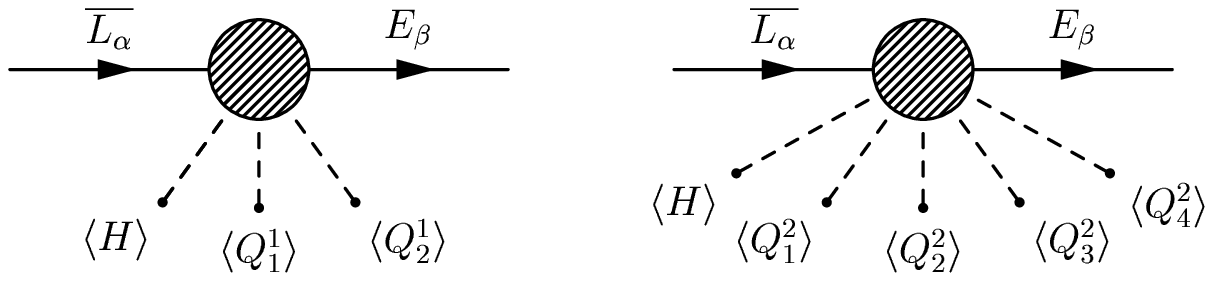}
\end{center}
\vspace{-3mm}
\caption{\small{Lowest-dimensional Yukawa interactions generating for
  $\alpha,\beta=\mu,\tau$ the charged lepton mass terms in the
  $\mu$-$\tau$-subsector.}}
  \label{fig:muontaumasses}
\end{figure}

Let us now suppose that the charges of the product gauge groups
$G_L\times G_R$, $G^1$, and $G^2$ are approximately conserved in the charged
lepton sector, implying the generation of hierarchical charged lepton mass terms via
the Froggatt-Nielsen mechanism. Furthermore, we assume that the fields
$\xi_0,\xi_1,$ and $\xi_2$ carry nonzero $G^3$ and $G^4$ quantum
numbers\footnote{The exact $G^3$ and $G^4$ charge assignment
to these fields is discussed in Sec.~\ref{sec:orbifold}.}, respectively,
while the fundamental charged Froggatt-Nielsen messengers transform only
trivially under $G^3$ and $G^4$. As a consequence, the fields $\xi_0,\xi_1,$ and
$\xi_2$ can be discarded in the discussion of the generation of the charged
lepton masses (see also Sec.~\ref{sec:orbifold}). Then, we conclude from table
\ref{tab:GLGRcharges} that gauge-invariance under
the product gauge group $G_L\times G_R\times G^1\times G^2$ allows
in the $\mu$-$\tau$-subsector only
two general types of non-renormalizable leading-order charged lepton mass
terms: One operator of the dimension $4+N_1=6$ and one of the
dimension $4+N_2=8$. The corresponding Froggatt-Nielsen-type diagrams are shown
in Fig.~\ref{fig:muontaumasses}. Note
here, that the effective Yukawa couplings of the dimension-six and
dimension-eight operators arise from the link fields of the deconstructed
extra-dimensional gauge groups $G^1$ and $G^2$, respectively. 

\subsection{Discrete charges}\label{sec:discretecharges}
It has been pointed out, that a naturally maximal $\nu_\mu$-$\nu_\tau$-mixing
is a strong hint for an underlying non-Abelian flavor symmetry acting on
the 2nd and 3rd generations \cite{moha99,wett99,king01}.
Seemingly, this flavor symmetry is unlikely to be a
continuous global symmetry because of the absence of such symmetries in
string theory \cite{banks88}. From the point of view of string theory, however, it is
interesting to analyze conventional grand unification in connection with {\it
discrete} symmetries since these might provide a solution to the doublet-triplet
splitting problem \cite{barb94,barr97,witt02}. Additionally, discrete symmetries
can ensure an approximately flat potential for the GUT breaking fields in order
to avoid fast proton decay \cite{dine02}. Moreover, it has recently been
demonstrated that the appropriate discrete symmetries of the low-energy theory
can be naturally obtained in deconstructed models with very coarse
latticization, when the lattice link variables are themselves subject to a
discrete symmetry \cite{witt02}. It is interesting to note, that
although a classical lattice gauge theory based on a system of link
fields which are elements of some discrete group has no continuum limit, this
does not necessarily carry over to the quantum theory\footnote{For a pedagogical
introduction to lattice gauge theory see, {\it e.g.}, Ref.~\cite{creu83}.}.

Among the discrete symmetries that have been proposed in the context of the MSW
LMA solution as a possible origin of an approximately maximal atmospheric mixing
angle $\theta_{23}$ are the symmetric groups $S_2$ and $S_3$ acting on the 2nd
and 3rd generations of leptons \cite{moha000,grim001}. While this can indeed give an approximately
maximal $\nu_\mu$-$\nu_\tau$-mixing, the hierarchical charged lepton mass
spectrum is typically not produced in these types of models, since they 
rather yield masses of the muon and tau that are of the same order of magnitude.
However, when we add to the regular representation of $S_2$ in the
$\mu$-$\tau$-flavor basis the generator
${\rm diag}(-1,\:1)$, one obtains the vector representation of the dihedral
group $\mathscr{D}_4$, which is non-Abelian (see App. \ref{app:dihedral}).
Since the generator ${\rm diag}(-1,\:1)$ distinguishes between $L_\mu$ and
$L_\tau$ it may serve as a possible source for the charged lepton mass
hierarchy, which breaks the permutation symmetry $S_2\subset \mathscr{D}_4$ characteristic
for the $\nu_\mu$-$\nu_\tau$-sector. Clearly, if the charged 
lepton masses arise from the Froggatt-Nielsen
mechanism, we will have to expect that the underlying symmetry is actually
equivalent to a group {\it extension}\footnote{An {\it extension} of a group $N$ by a group
$H$ is an embedding of $N$ into some group $E$ such that $N$ is normal in $E$
and $H\simeq E/N$.} of some subgroup of $\mathscr{D}_4$,
presumably a suitable subgroup of some replicated product of $\mathscr{D}_4$-factors.

Motivated by these general observations, we will take here the view, that an approximately maximal atmospheric mixing
angle $|\theta_{23}-\frac{\pi}{4}|\ll 1$ is due to a non-Abelian discrete
symmetry $\mathscr{G}$ which is a group
extension based on the generators of the dihedral group $\mathscr{D}_4$
under which the leptons of the 2nd and 3rd generation, the scalars
$\xi_1,\xi_2,$ and the
scalar link variables of the deconstructed extra-dimensional gauge symmetries
$G^1=\Pi^2_{i=1}G^1_i$ and $G^2=\Pi^4_{i=1}G^2_i$ transform non-trivially.
Specifically, we suppose that in the
defining representation $\mathcal{D}$ the group
$\mathscr{G}$, which is a subgroup of the four-fold (external) product group
$\mathscr{D}_4\times\mathscr{D}_4\times\mathscr{D}_4\times\mathscr{D}_4$,
can be written as a sequence
\begin{equation}\label{eq:subreps}
\mathcal{D}(g)\equiv {\rm diag}\left[
\begin{matrix}
 D_1(g), & D_2(g), & D_3(g), & D_4(g)
\end{matrix}
\right]\quad g\in\mathscr{G},  
\end{equation}
where each element $g\in\mathscr{G}$ is associated with four\footnote{Since
each element $g\in{\mathscr{G}}$ is associated with
four (in general) different operators
$D_1(g),D_2(g),D_3(g),$ and $D_4(g)$,
one may view in a discretized picture $\mathcal{D}$ as a
4-valued representation of $\mathscr{D}_4$ with
$\mathscr{G}$ as the corresponding (universal) covering group.}
(in general different) $2\times 2$ charge operators $D_i(g)$
$(i=1,2,3,4)$ which have values in the vector representation of $\mathscr{D}_4$.
Clearly, for given $i=1,2,3,4$ the set of all operators
$\{D_i(g):g\in\mathscr{G}\}$
forms a subrepresentation of $\mathscr{G}$ which we will call
$D_i$. Now, using the notation of App. \ref{app:dihedral} we suppose
that $\mathscr{G}$ is represented by four generators with following
$\mathscr{D}_4$-charge structure
\begin{subequations}\label{eq:abstractgenerators}
\begin{eqnarray}
 \mathcal{D}(a_1)&\equiv&
 {\rm diag}
 \left[\begin{matrix} 
  D(C_b), & D(E), & D(C_{b'}), & D(C_{b'})\end{matrix}\right],\\
 &&\nonumber\\
 \mathcal{D}(a_2)&\equiv&
 {\rm diag}
 \left[\begin{matrix}
  D(C_{b'}), & D(C_{b'}), & D(C_b), & D(E) \end{matrix}\right],\\
 &&\nonumber\\
 \mathcal{D}(a_3)&\equiv&
 {\rm diag}
 \left[\begin{matrix}
  D(E), & D(E), & D(C_b), & D(C_b)\end{matrix}\right],\\
 &&\nonumber\\
 \mathcal{D}(a_4)&\equiv&
 {\rm diag}
 \left[
 \begin{matrix}
 D(C_b), & D(C_b), & D(E), & D(E)
 \end{matrix}
 \right],
\end{eqnarray}
\end{subequations}
where $a_1,a_2,a_3,a_4\in\mathscr{G}$ are the corresponding abstract
generators. Note that these operators are characterized by a
one-to-one-correspondence  $\mathcal{D}(a_1)\leftrightarrow\mathcal{D}(a_2)$
and $\mathcal{D}(a_3)\leftrightarrow\mathcal{D}(a_4)$ under the permutation of the
upper-left and the lower-right $4\times 4$-matrices.
As will be shown in Sec. \ref{sec:G}, by factoring the subrepresentation
$D_i$ for any $i=1,2,3,4$ with respect to its kernel $\mathcal{N}_i$
we obtain a representation of the factor group $\mathscr{G}/\mathcal{N}_i$ 
that is isomorphic with $\mathscr{D}_4$. This implies, of course, that all four
subrepresentations $D_1,D_2,D_3,$ and
$D_4$ are two-dimensional irreducible representations (irreps) of
$\mathscr{G}$. From App. \ref{app:dihedral} it is seen, that in an 
appropriate basis the generators of Eqs.~(\ref{eq:abstractgenerators}) can be
explicitly written as
\begin{subequations}\label{eq:generators}
\begin{eqnarray}
 \mathcal{D}(a_1)&\equiv&
 {\rm diag}
 \left[
 \begin{matrix}
 \left(
 \begin{matrix}
  -1&0\\
  0&1
 \end{matrix}
 \right), &
 \left(
 \begin{matrix}
  1&0\\
  0&1
 \end{matrix}
 \right)
 , &
 \left(
 \begin{matrix}
  0&1\\
  1&0
 \end{matrix}
 \right),&
 \left(
 \begin{matrix}
  0&1\\
  1&0
 \end{matrix}
 \right)
 \end{matrix}
 \right],\\
 &&\nonumber\\
 \mathcal{D}(a_2)&\equiv&
 {\rm diag}
 \left[
 \begin{matrix}
 \left(
 \begin{matrix}
  0&1\\
  1&0
 \end{matrix}
 \right), &
  \left(
 \begin{matrix}
  0&1\\
  1&0
 \end{matrix}
 \right), &
 \left(
 \begin{matrix}
  -1&0\\
  0&1
 \end{matrix}
 \right),
 &
 \left(
 \begin{matrix}
  1 &0\\
  0&1
 \end{matrix}
 \right)
 \end{matrix}
 \right],\\
 &&\nonumber\\
 \mathcal{D}(a_3)&\equiv&
 {\rm diag}
 \left[
 \begin{matrix}
 \left(
 \begin{matrix}
  1&0\\
  0&1
 \end{matrix}
 \right), &
  \left(
 \begin{matrix}
  1&0\\
  0&1
 \end{matrix}
 \right)
 ,&
 \left(
 \begin{matrix}
  -1&0\\
  0&1
 \end{matrix}
 \right),
 &
 \left(
 \begin{matrix}
  -1&0\\
  0&1
 \end{matrix}
 \right)
 \end{matrix}
 \right],\\
 &&\nonumber\\
 \mathcal{D}(a_4)&\equiv&
 {\rm diag}
 \left[
 \begin{matrix}
 \left(
 \begin{matrix}
  -1&0\\
  0&1
 \end{matrix}
 \right), &
  \left(
 \begin{matrix}
  -1&0\\
  0&1
 \end{matrix}
 \right)
 ,&
 \left(
 \begin{matrix}
  1&0\\
  0&1
 \end{matrix}
 \right),
 &
 \left(
 \begin{matrix}
  1&0\\
  0&1
 \end{matrix}
 \right)
 \end{matrix}
 \right].
\end{eqnarray}
\end{subequations}
With respect to $\mathcal{D}$ we will combine the left- and right-handed SM
leptons as well as the Dirac neutrinos of the 2nd and 3rd generations into the 
$\mathscr{G}$-doublets $\mathbf{2}_\ell\equiv(L_\mu,\:L_\tau)^T$,
$\mathbf{2}_E\equiv (E_\mu,E_\tau)^T$, and
$\mathbf{2}_F\equiv(F_\mu,\:F_\tau)^T$, respectively. The fermionic doublets
$\mathbf{2}_\ell$ and $\mathbf{2}_F$ as well as the scalars
$\xi_1\equiv\left(\xi_{1a},\:\xi_{1b}\right)^T$ and
$\xi_2\equiv\left(\xi_{2a},\:\xi_{2b}\right)^T$ are all put into the doublet representation $D_1$.
Next, the generalized Wigner-Eckart theorem tells us that the effective Yukawa
interaction matrix spanned by $\mathbf{2}_\ell$ and $\mathbf{2}_E$
in the $\mu$-$\tau$-subsector of the charged leptons is identical
with a linear combination of sets of irreducible Yukawa tensor operators.
If the irreducible Yukawa tensor operators take their values in the vector representation of $\mathscr{D}_4$, it
is clear from App.~\ref{app:dihedral}, that the hierarchy
$m_\mu\ll m_\tau$ is only possible if $\mathbf{2}_E$ transforms according to an
irrep of $\mathscr{G}$ which is inequivalent\footnote{Similarity transformations
allow only mappings within one class, which would yield $m_\mu=m_\tau$
after SSB. Note also that different transformation properties of
left-handed and right-handed fermions under horizontal symmetries are used in
models of ``neutrino democracy'' \cite{shafi02}.} with $D_1$. We will therefore put
$\mathbf{2}_E$ into the irrep $D_2$ which is inequivalent with $D_1$
and note that the large subgroup of $\mathscr{G}$ generated by
$\mathcal{D}(a_2),\mathcal{D}(a_3),$ and $\mathcal{D}(a_4)$ acts diagonally on
$\mathbf{2}_\ell$ and $\mathbf{2}_E$.

\begin{figure}
\begin{center}
\includegraphics*[bb = 151 581 432 675]{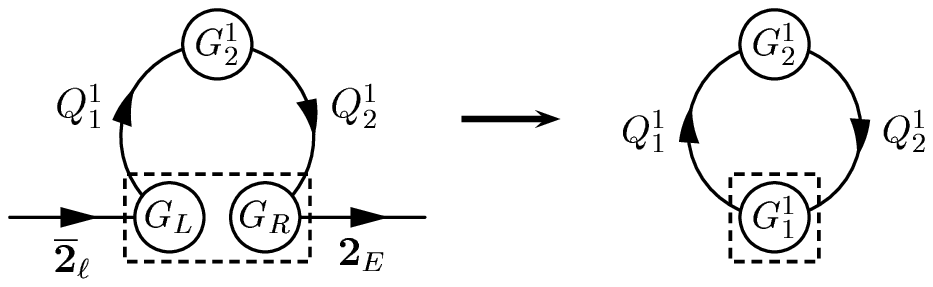}
\end{center}
\vspace{-3mm}
\caption{\small{Contour in theory space which generates the dimension-six mass
  operator in Fig.~\ref{fig:muontaumasses}
  by connecting $\mathbf{2}_\ell$ and $\mathbf{2}_E$ via the link fields of $G^1$
  (left panel). Formal contraction of the open fermion lines and
  $G_L\times G_R$ into $G^1_1$ (dashed boxes)
  exhibits the full $G^1$ gauge-invariance of the corresponding
  effective Yukawa couplings (right panel).}}
  \label{fig:contour1}
\end{figure}

\begin{figure}
\begin{center}
\includegraphics*[bb = 146 563 440 682]{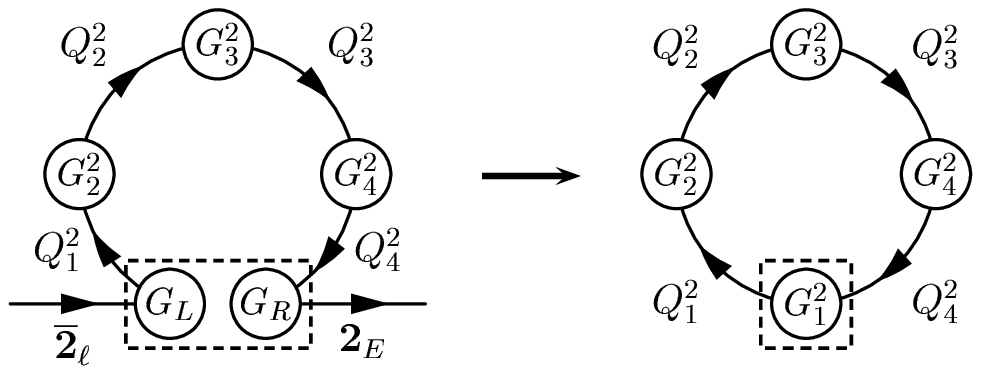}
\end{center}
\vspace{-3mm}
\caption{\small{Contour in theory space which generates the dimension-eight
  mass operator in Fig.~\ref{fig:muontaumasses}
  by connecting $\mathbf{2}_\ell$ and $\mathbf{2}_E$ via
  the link fields of $G^2$ (left panel). Formal contraction of the open fermion lines
  and $G_L\times G_R$ into $G^2_1$ (dashed boxes)
  exhibits the full $G^2$ gauge-invariance of the corresponding
  effective Yukawa couplings (right panel).}}
  \label{fig:contour2}
\end{figure}

In Sec.~\ref{sec:U(1)charges} we have seen that
$G_L\times G_R\times G^1\times G^2$ gauge-invariance requires the possible
lowest-dimensional
effective Yukawa interactions between $\mathbf{2}_\ell$ and $\mathbf{2}_E$
to be identical with the dimension-six and dimension-eight
Froggatt-Nielsen-type operators displayed in Fig. \ref{fig:muontaumasses}.
In theory space, we can therefore identify the sets of irreducible Yukawa tensor
operators spanned by $\mathbf{2}_\ell$ and $\mathbf{2}_E$ with the
Wilson loops around the plaquettes associated with the deconstructed
extra-dimensional gauge-symmetries $G^1$ and $G^2$:
The dimension-six and dimension-eight operators correspond to the Wilson loops
around the moose diagrams of $G^1$ (Fig.~\ref{fig:contour1}) and $G^2$
(Fig.~\ref{fig:contour2}), respectively.
Although the irreps $D_1$ and $D_2$ already determine
the overall transformation properties of the Wilson loops under $\mathscr{G}$,
there is still some ambiguity in the individual $\mathscr{G}$-charge assignment
to the involved link-fields $Q^1_1,Q^1_2,$ and 
$Q^2_{1},Q^2_{2},Q^2_{3},Q^2_4$.
Here, it is appealing to assume that all the scalar link fields of $G^1$
and $G^2$ transform according to some doublet subrepresentation of
$\mathscr{G}$, which has the property that its lift is isomorphic 
with $\mathscr{D}_4$. Furthermore, a non-trivial Yukawa matrix structure
requires that the products
$\Pi_{i=1}^2 Q^1_{i}$ and $\Pi_{i=1}^4 Q^2_i$
are non-singlet representations of $\mathscr{G}$. Since each of the
Wilson loops involves an even number of (two or four) link fields, we find from the
multiplication rules in App.~\ref{app:dihedral} that this can only be the case
if each of the sets $\{Q^1_1,Q^1_2\}$ and $\{Q^2_1,Q^2_2,Q^2_3,Q^2_4\}$
contains at least two inequivalent irreps. This can be simply realized by
putting, {\it e.g.}, the
1st ($i=1$) link fields $Q^1_1$ and $Q^2_1$ into the doublet representation
$D_3$ while the remaining link fields 
$Q^1_2,Q^2_2,Q^2_3,$ and $Q^2_4$ all transform as doublets under
$D_4$. Note again, that both $D_3$ and $D_4$ are characterized by a large
common subgroup generated by
$\mathcal{D}(a_1),\mathcal{D}(a_3),$ and $\mathcal{D}(a_4),$ which acts
diagonally on all of the link fields. In component-form, the
scalar ${\mathscr{G}}$-doublets will be written as
$Q^1_i\equiv\left(q^1_{ia},\:q^1_{ib}\right)^T$, where $i=1,2,$ and
$Q^2_i\equiv\left(q^2_{ia},\:q^2_{ib}\right)^T$, where $i=1,2,3,4$.
To complete the $\mathscr{G}$-charge assignment, we suppose that the first
generation fields $L_e,E_e,$ and $F_e$, as well as the scalar $\xi_0$ transform
only trivially under $\mathscr{G}$, {\it i.e.}, they are all put into the
identity representation $I$ of $\mathscr{G}$. For the fermionic
$\mathscr{G}$-singlets $L_e,E_e,$ and $F_e$ we will also choose the notation
$\mathbf{1}_e\equiv L_e$, $\mathbf{1}_E\equiv E_e$, and
$\mathbf{1}_F\equiv F_e$.  The assignment of the fields to the irreps $D_i$
$(i=1,2,3,4)$ and $I$ is summarized in table \ref{tab:subreps}.

\begin{table}
\begin{center}
\begin{tabular}{|c|c|c|c|c|}
\hline
     $I$ &$D_1$ & $D_2$ & $D_3$ &
	 $D_4$\\ \hline \hline
     $\mathbf{1}_e,\mathbf{1}_E,\mathbf{1}_F,\xi_0$ &
	 $\mathbf{2}_\ell,\mathbf{2}_F,\xi_1,\xi_2$ &
	 $\mathbf{2}_E$ & $Q^1_1,Q^2_1$ &
	$Q^1_2,Q^2_2,Q^2_3,Q^2_4$\\
\hline
\end{tabular}
\end{center}
\caption{Assignment of the fermionic and scalar fields to the different
irreps $D_i$ and $I$. Note that the scalar link fields transform according to
their position in the ``index space'' of gauge groups.}\label{tab:subreps}
\end{table}

In this subsection, we have presented the non-Abelian group $\mathscr{G}$ in
terms of its generators which are motivated by phenomenology.
In the following subsection we will examine in some more detail the
structure of $\mathscr{G}$ through a descending series of normal
subgroups, {\it i.e.}, the normal structure.

\subsection{Normal structure}\label{sec:G}
In Sec.~\ref{sec:discretecharges} the discrete group $\mathscr{G}$ has been
constructed from the generators of the dihedral group $\mathscr{D}_4$. A standard way of gaining
further information about $\mathscr{G}$ is to analyze a series of subgroups of
$\mathscr{G}$, where each term is either normal in $\mathscr{G}$ or at least
normal in the previous term. In general, if a subgroup $N$ of some group $G$ is
normal in $G$ we will write $N\trianglelefteq G$.

We will denote by ${\mathcal{K}}_1$ the collection of all elements
$g\in \mathscr{G}$ which obey $D_1(g)=\mathbbm{1}_2$, where $\mathbbm{1}_2$
is the $2\times 2$ unit matrix. Accordingly, we will define
${\mathcal{K}}_2$ as the set of all elements $g\in \mathscr{G}$ which
obey $D_1(g)=D_2(g)=\mathbbm{1}_2$, and ${\mathcal{K}}_3$ as the set of all
elements $g\in\mathscr{G}$ which obey $D_1(g)=D_2(g)=D_3(g)=\mathbbm{1}_2$.
We therefore have the sequence
\begin{equation}\label{eq:subgroupseries}
{\mathcal{K}}_3\subset {\mathcal{K}}_2 \subset {\mathcal{K}}_1
\subset\mathscr{G},
\end{equation}
where each subset is a group, actually an invariant subgroup of the embedding
groups, for if $a\in {\mathcal{K}}_i$ $(i=1,2,3)$ is homomorphically mapped on
the identity of the operator group $D_1\oplus\ldots\oplus D_{i}$, then so are
all elements in its class. We therefore have
$\mathcal{K}_{i+1}\trianglelefteq\mathcal{K}_i$ and
$\mathcal{K}_i\trianglelefteq\mathscr{G}$ for every $i$, {\it i.e.},
the subgroup series in Eq.~(\ref{eq:subgroupseries}) is in fact a
{\it normal series}.
From Eqs.~(\ref{eq:generators}) we find that for any
$k\in{\mathcal{K}}_2$ the decomposition of $\mathcal{D}(k)$ into a product of
the operators $\mathcal{D}(a_i)$ $(i=1,2,3,4)$ necessarily involves
each of the factors $\mathcal{D}(a_1),\mathcal{D}(a_2),$ and
$\mathcal{D}(a_4)$ an even number of times. In turn, this implies that the
operators $D_3(k)$ and $D_4(k)$ are on diagonal form
and can take their values only in the classes $E,C_4^2,$ and $C_2(2)$ of the
dihedral group $\mathscr{D}_4$ (see App.~\ref{app:dihedral}).
Since the number of elements in the classes $E,C^2_4,$ and $C_2(2)$ is four,
we conclude that the order of ${\mathcal{K}}_2$, which we will denote by
$|\mathcal{K}_2|$, obeys $|\mathcal{K}_2|\leq 4$.
Indeed, besides the unity, we find the three distinct elements $a_3$,
$a_1a_3a_1$, and $(a_1a_3)^2$ which are all contained in ${\mathcal{K}}_2$.
Hence, $|\mathcal{K}_2|=4$ and we have
\begin{eqnarray}\label{eq:K_2}
 {\mathcal{K}}_2&\simeq&\left\{
 {\rm diag}\left[D(E),D(E),D(E),D(E)\right],
 \right.
 \nonumber\\
 \nonumber\\
 &&
 \:\:
 {\rm diag}\left[D(E),D(E),D(C_a),D(C_a)\right],
 \nonumber\\
 \nonumber\\
&&
 \:\:
 {\rm diag}\left[D(E),D(E),D(C_b),D(C_b)\right],
 \nonumber\\
 \nonumber\\
 &&
 \:\:
 \left.
 {\rm diag}\left[D(E),D(E),D(C_4^2),D(C_4^2)\right]
 \right\}.
\end{eqnarray}
Since
${\mathcal{K}}_3\subset{\mathcal{K}}_2$ the group ${\mathcal{K}}_3$ is trivial and
contains only the identity of $\mathcal{D}$, as inspection of Eq.~(\ref{eq:K_2}) shows.
From Eq.~(\ref{eq:K_2}) we find that $\mathcal{K}_2$ describes a 2-fold axis
with a system of two 2-fold axes at right angle to it. Hence, $\mathcal{K}_2$ is
isomorphic with the {\it Klein group} $Z_2\times Z_2\simeq \mathcal{K}_2$.
In fact, the Klein group is one of the dihedral groups
$Z_2\times Z_2\simeq\mathscr{D}_2$ (see App.~\ref{app:dihedral}).
With ${\mathcal{K}}_2$ explicitly given in Eq.~(\ref{eq:K_2})
we can now easily construct a representation of the embedding group
${\mathcal{K}}_1\trianglerighteq \mathcal{K}_2$. First,
note that for any $k\in{\mathcal{K}}_1$ the resolution of the operator
$\mathcal{D}(k)$ into a
product of the generators in Eq.~(\ref{eq:generators}) must involve
$\mathcal{D}(a_2)$ an even number of times. Hence, $D_2(k)$ is necessarily
on diagonal form
implying that the index of the subgroup ${\mathcal{K}}_2$ under the group
${\mathcal{K}}_1$ is at most four. In fact, the group ${\mathcal{K}}_1$ can be 
decomposed into (right) cosets\footnote{Since the subgroup ${\mathcal{K}}_2$ is
invariant, left and right cosets are identical.} in terms of
\begin{equation}\label{eq:K_1cosets}
{\mathcal{K}}_1\simeq{\mathcal{K}}_2+\sum_{i=1}^3{\mathcal{K}}_2\:b_i,
\end{equation}
where we can choose the different coset representatives to be
$b_1=a_1a_4$, $b_2=a_2b_1a_2$, and $b_3=b_2b_1$. In other
words, four elements of ${\mathcal{K}}_1$ are mapped on each element
of the operator group corresponding to the representation of $\mathcal{K}_1$
subduced by $D_2$ (subduced representation)
$D_2\downarrow\mathcal{K}_1$. Since the subgroup $\mathcal{K}_2\trianglelefteq
\mathcal{K}_1$ is the kernel of $D_2\downarrow\mathcal{K}_1$, the
operators $D_2(b_i)$ $(i=1,2,3)$ from the (right) transversal allow us to
identify the lift of $D_2\downarrow \mathcal{K}_1$ as the Klein group
$Z_2\times Z_2$. As for the coset representatives don't commute with the
elements of ${\mathcal{K}}_2$ it follows
that ${\mathcal{K}}_1$ is actually an (external)
{\it semi-direct} product or {\it split extension} of $\mathcal{K}_2$ by
$Z_2\times Z_2$. Hence, we can write
\begin{equation}
{\mathcal{K}}_1\simeq 
(Z_2\times Z_2)\rtimes_\varphi (Z_2\times Z_2),
\end{equation}
where $\varphi$ is the mapping of $\mathcal{K}_2$ into the automorphism group
of $\mathcal{K}_2$ which is determined by the particular choice of the generators
in Eq.~(\ref{eq:abstractgenerators}). The mapping $\varphi$ can be
identified with the conjugation homomorphism of the corresponding internal
semi-direct product describing the interaction of
the two involved $Z_2\times Z_2$ subgroups inside $\mathscr{G}$.
Now, the order of ${\mathcal{K}}_1$ is simply given by the product of the
orders of the semi-direct factors, {\it i.e.}, $|\mathcal{K}_1|=16$.
Since we have $\mathscr{G}/{\mathcal{K}}_1\simeq \mathscr{D}_4$ the decomposition of
$\mathscr{G}$ into right
cosets with respect to ${\mathcal{K}}_1$ reads
\begin{equation}
 \mathscr{G}\simeq{\mathcal{K}}_1+\sum_{i=1}^7{\mathcal{K}}_1\:c_i,
\end{equation}
where we can choose for the coset representatives $c_1=a_1,$ $c_2=a_2a_1a_2$,
$c_3=(a_1a_2)^2$, $c_4=a_2$, $c_5=a_1a_2$, $c_6=a_2a_1$, and $c_7=a_1a_2a_1$.
Here, the irrep $D_1$ maps 16 elements of $\mathscr{G}$ onto each element of
$\mathscr{D}_4$ and hence $|\mathscr{G}|=128$. Again, the seven coset
representatives don't commute with the
elements of ${\mathcal{K}}_1$ which shows that $\mathscr{G}$ is a split
extension of $\mathcal{K}_1$ by $\mathscr{D}_4$, {\it i.e.},
\begin{equation}\label{eq:G}
 \mathscr{G}\simeq (Z_2\times Z_2)\rtimes_\varphi
 (Z_2\times Z_2)\rtimes_\psi\mathscr{D}_4,
\end{equation}
where $\psi$ is identified with the conjugation homomorphism describing the
interaction of $\mathscr{D}_4$ with the involved $Z_2\times Z_2$
subgroups inside $\mathscr{G}$. By construction, the operator group associated
with the irrep $D_1$ has $\mathcal{K}_1$ as its kernel. Hence, by factoring
$D_1$ with respect to the subgroup $\mathcal{K}_1$ we see that the
lift of $D_1$ is isomorphic with $\mathscr{D}_4$. Furthermore,
by replacing $\mathcal{D}(a_1)\rightarrow\mathcal{D}(a_1)\mathcal{D}(a_4)$, we
confirm that the irreps $D_1$ and $D_2$ change their r\^oles in the
above considerations. Taking, in addition, the exchange symmetry
between $\mathcal{D}(a_1)\leftrightarrow\mathcal{D}(a_2)$ and
$\mathcal{D}(a_3)\leftrightarrow\mathcal{D}(a_4)$ under permutation of the
$4\times 4$ submatrices on the diagonal into account
(see Sec. \ref{sec:discretecharges}), one generally concludes that the lift of
every subrepresentation $D_i$ $(i=1,2,3,4)$ is isomorphic with
$\mathscr{D}_4$.

In this subsection we have analyzed the structure of the group $\mathscr{G}$ and
its relation to the dihedral groups $\mathscr{D}_2$ and $\mathscr{D}_4$.
Specifically, we have seen that the lift of any irrep $D_i$ $(i=1,2,3,4)$ is
isomorphic with $\mathscr{D}_4$. This will allow us in the next section to apply
the decomposition and multiplication rules of $\mathscr{D}_4$ in order to
find the relevant $\mathscr{G}$-singlets in product representations.

\section{Construction of the scalar potential}\label{sec:scalarpotential}
We will denote by $\Phi$ and $\Omega$ two arbitrary scalar
${\mathscr{G}}$-doublets which are listed in table \ref{tab:subreps} and write them in
component-form as
\begin{equation}\label{eq:doublets}
 \Phi\equiv
 \left(\phi_a,\:\phi_b\right)^T,\quad
 \Omega\equiv
 \left(\omega_a,\:\omega_b\right)^T,
\end{equation}
where $\Phi,\Omega\in\{\xi_1,\xi_2,Q^1_1,Q^1_2,Q^2_1,Q^2_2,Q^2_3,Q^2_4\}$. For
definiteness, we will assume that $\Phi$ is put into the irrep $D_i$
and $\Omega$ is put into the irrep $D_j$, where $i,j=1,3,4$.
Gauge-invariance under the $U(1)$ product groups $\Pi_{j=1}^4 G^j$ and
$G_L\times G_R$ tells us that any renormalizable term in the scalar potential
which involves, {\it e.g.}, the field $\Phi$ must
actually contain the tensor product $\Phi\otimes\Phi^\dagger$. The lowest-dimensional
$\mathscr{G}$-invariant operator-products of $\Phi$ in the multi scalar potential are therefore
\begin{equation}\label{eq:V_0}
 V_0(\Phi)\:\equiv\:\Phi\otimes\Phi^\dagger|_{1_A}=
 A_0\left(|\phi_a|^2+|\phi_b|^2\right),
\end{equation}
where $\Phi$ transforms under any of the irreps $D_i$ $(i=1,3,4)$ and $A_0$ is a
real-valued number.
Note that three-fold products of $\Phi$ and $\Phi^\dagger$ in the potential are
forbidden by both the $U(1)$ charge assignment as well as by
$\mathscr{G}$-invariance.
Accordingly, any invariant term in the scalar potential which mixes the doublets
$\Phi$ and $\Omega$ must be contained in the product representation
$\Phi\otimes\Phi^\dagger\otimes\Omega\otimes\Omega^\dagger$. Since the liftings
of the irreps $D_i$ and $D_j$ are isomorphic with $\mathscr{D}_4$ (see
Sec.~\ref{sec:G}), the
$\mathscr{G}$-invariant mixed terms of
$\Phi$ and $\Omega$ are found by considering all combinations of the
one-dimensional irreps of $\mathscr{D}_4$ in the product
\begin{equation}\label{eq:tensorproduct}
\Phi\otimes\Phi^\dagger\otimes\Omega\otimes\Omega^\dagger=
\left[\mathbf{1}_{\mathbf{A}}^i+\mathbf{1}_{\mathbf{B}}^i+
      \mathbf{1}_{\mathbf{C}}^i+\mathbf{1}_{\mathbf{D}}^i\right]\otimes
\left[\mathbf{1}^j_{\mathbf{A}}+\mathbf{1}_{\mathbf{B}}^j+
      \mathbf{1}_{\mathbf{C}}^j+\mathbf{1}_{\mathbf{D}}^j\right],
\end{equation}
where each of the sets 
$\mathbf{1}_{\mathbf{A}}^i$,$\mathbf{1}_{\mathbf{B}}^i$,
$\mathbf{1}_{\mathbf{C}}^i$, $\mathbf{1}_{\mathbf{D}}^i$, and
$\mathbf{1}_{\mathbf{A}}^j$,$\mathbf{1}_{\mathbf{B}}^j$,
$\mathbf{1}_{\mathbf{C}}^j$,$\mathbf{1}_{\mathbf{D}}^j$ respectively denotes
the $\mathscr{D}_4$-singlet representations associated with
$D_i$ and $D_j$ 
(see App. \ref{app:dihedral}). In component form, the singlet representations
are explicitly given in Eq.~(\ref{eq:singlets}). 
In order to extract from Eq.~(\ref{eq:tensorproduct}) the relevant
dimension-four terms we will first suppose
that $\Phi$ and $\Omega$ are put into the same irrep $D_i=D_j$.
Then, the
decomposition of the product representations in Eq.~(\ref{eq:2x2}) in
conjunction with the multiplication rules in table \ref{tab:multiplication} yield that
the invariant mixed terms of $\Phi$ and $\Omega$ are in this case
\begin{subequations}\label{eq:V_123}
\begin{eqnarray}
 V_1(\Phi,\Omega)&\equiv&(\Phi\otimes\Phi^\dagger)\otimes(\Omega\otimes\Omega^\dagger)|_{1_A}
       \nonumber\\
 &=& A_1\left(|\phi_a|^2+|\phi_b|^2\right)
      \left(|\omega_a|^2+|\omega_b|^2\right)\nonumber\\
 && +B_1\left(\phi_a\phi_b^\dagger-\phi_a^\dagger\phi_b\right)
     \left(\omega_a\omega_b^\dagger-\omega_a^\dagger\omega_b\right)
	 \nonumber\\
 &&+C_1\left(|\phi_a|^2-|\phi_b|^2\right)\left(|\omega_a|^2-|\omega_b|^2\right)
      \nonumber\\
 &&+D_1\left(\phi_a\phi_b^\dagger+\phi_a^\dagger\phi_b\right)
     \left(\omega_a\omega_b^\dagger+\omega_a^\dagger\omega_b\right),
	 \label{eq:V_1}
\end{eqnarray}
where we have labeled the real-valued coefficients $A_1,B_1,C_1,$ and $D_1$
of the invariants according to the sequence of products
$(\mathbf{1_A})^2,(\mathbf{1_B})^2,(\mathbf{1_C})^2,$ and
$(\mathbf{1_D})^2$.
Let us now turn to the case, when $\Phi$ and $\Omega$ belong to different
irreps $D_i\neq D_j$.
First, suppose that $\Phi$ transforms under $D_3$ and $\Omega$
transforms under $D_4$. Application of the generators
$\mathcal{D}(a_1),\mathcal{D}(a_2),$ and $\mathcal{D}(a_3)$ yields that in
Eq.~(\ref{eq:tensorproduct}) for $i=3$ and $j=4$ only the combinations
$\mathbf{1}_{\mathbf{A}}^i\otimes\mathbf{1}_{\mathbf{A}}^j$ and
$\mathbf{1}_{\mathbf{C}}^i\otimes\mathbf{1}_{\mathbf{C}}^j$ are
$\mathscr{G}$-invariants.
Hence, the most general invariant mixed term of $\Phi$ and $\Omega$ reads
\begin{eqnarray}
 V_2(\Phi,\Omega)
 &\equiv& A_2\left(|\phi_a|^2+|\phi_b|^2\right)
      \left(|\omega_a|^2+|\omega_b|^2\right)\nonumber\\
 &&+C_2\left(|\phi_a|^2-|\phi_b|^2\right)\left(|\omega_a|^2-|\omega_b|^2\right),
	 \label{eq:V_2}
\end{eqnarray}
where $\Phi\in\{Q^1_1,Q^2_1\}$, $\Omega\in\{Q^1_2,Q^2_2,Q^2_3,Q^2_4\}$ and
the coefficients $A_2,C_2$ are real-valued numbers. Now, suppose that $\Phi$ transforms under $D_1$ and $\Omega$
transforms under one of the irreps $D_3$ or $D_4$. In this case, subsequent
application of the operators $\mathcal{D}(a_i)$ $(i=1,2,3,4)$ to
Eq.~(\ref{eq:tensorproduct}) readily yields that the most general mixed term of the
fields $\Phi$ and $\Omega$ is given by
$\mathbf{1}_{\mathbf{A}}^i\otimes\mathbf{1}_{\mathbf{A}}^j$ and hence
\begin{equation}\label{eq:V_3}
 V_3(\Phi,\Omega)\:\equiv\:
 A_3\left(|\phi_a|^2+|\phi_b|^2\right)\left(|\omega_a|^2+|\omega_b|^2\right),
\end{equation}
\end{subequations}
where $\Phi\in\{\xi_1,\xi_2\}$,
$\Omega\in\{Q^1_1,Q^1_2,Q^2_1,Q^2_2,Q^2_3,Q^2_4\}$ and the coefficient $A_3$ is
a real-valued number.
Putting everything together, the most general multi scalar potential $V$ of the
SM singlet scalar fields decomposes into the terms $V_0(\Phi)$,$V_1(\Phi,\Omega)$,
$V_2(\Phi,\Omega)$, and $V_3(\Phi,\Omega)$ of Eqs.~(\ref{eq:V_123}), as follows
\begin{eqnarray}\label{eq:V}
 V&\equiv&\sum_{i,j,k}\left[V_0(\xi_i)+V_0(Q^1_j)+V_0(Q^2_k)\right]
 +\sum_{i_1,i_2=1,2}V_1(\xi_{i_1},\xi_{i_2})\nonumber\\
  &&+\sum_{X,Y=Q^1_1,Q^2_1}V_1(X,Y)
  +\sum_{j_1,j_2=2}^4\left[V_1(Q^1_2,Q^2_{j_1})+V_1(Q^2_{j_1},Q^2_{j_2})
  \right]\nonumber\\
  &&+V_1(Q^1_2,Q^1_2)+
  \sum_{X=Q^1_1,Q^2_1}\sum_{j_1=2}^4\left[
  V_2(X,Q^1_2)+V_2(X,Q^2_{j_1})\right]
  \nonumber\\
  &&+\sum_{i,j,k}\left[V_3(\xi_i,Q^1_j)+V_3(\xi_i,Q^2_k)\right],
\end{eqnarray}
where $i=1,2;\:j=1,2$, and $k=1,2,3,4$. In Eq.~(\ref{eq:V}) we have
omitted $\xi_0$ and the $SU(2)$ Higgs doublet $H$. Actually, in any
renormalizable terms of the full multi-scalar potential which mix the
$\mathscr{G}$-singlets with the $\mathscr{G}$-doublets, the
$\mathscr{G}$-singlet
fields $\xi_0$ and $H$ are only allowed to appear in terms of their
absolute squares $|\xi_0|^2$ and $|H|^2$. This is an immediate
consequence of the $G_L\times G_R$ charge of $\xi_0$ and the
electroweak quantum numbers of $H$. As a result, there
exists a range of parameters in the multi-scalar potential where the vacuum
alignment of the $\mathscr{G}$-doublet scalars is essentially independent
from the details of the VEVs $\langle\xi_0\rangle$ and $\langle H\rangle$.
Specifically, we can assume the standard electroweak symmetry breaking and
allow the field $\xi_0$ to aquire an arbitrary VEV of the order
$|\langle\xi_0\rangle|\simeq 10^2\:\rm{GeV}$. It is therefore sufficient to
restrict our considerations concerning the vacuum alignment of the
$\mathscr{G}$-doublet scalars to the potential $V$ in Eq.~(\ref{eq:V}),
in order to determine the range of parameters which leads to realistic
lepton masses and mixing angles. This analysis will be carried out in the
following section.

\section{The vacuum alignment mechanism}\label{sec:vacuumalignment}
We will now determine the vacuum structure emerging after SSB from the
potential $V$ by minimizing each of the individual potentials $V_1(\Phi,\Omega)$ and
$V_2(\Phi,\Omega)$ which appear in Eq.~(\ref{eq:V}). For this purpose it is
suitable to parameterize the VEVs of the doublets $\Phi$ and $\Omega$ as
\begin{subequations}\label{eq:parameterization}
\begin{eqnarray}
\langle\Phi\rangle&=&
 \left(
 \begin{matrix}
  \langle\phi_a\rangle\\
  \langle\phi_b\rangle
 \end{matrix}
 \right)\:\:=\:\:
 v_1
 \left(
 \begin{matrix}
  {\rm e}^{{\rm i}\varphi_1}\:{\rm cos}\:\alpha\\
  {\rm e}^{{\rm i}\varphi_1'}\:{\rm sin}\:\alpha
 \end{matrix}
 \right)
 \:\:\equiv\:\:
 v_1
 \left(
 \begin{matrix}
  {\rm e}^{{\rm i}\varphi_1}\:c_\alpha\\
  {\rm e}^{{\rm i}\varphi_1'}\:s_\alpha
 \end{matrix}
 \right),\\
 &&\nonumber\\
 \langle\Omega\rangle
 &=&
 \left(
 \begin{matrix}
  \langle\omega_a\rangle\\
  \langle\omega_b\rangle
 \end{matrix}
 \right)\:\:=\:\:
 v_2
 \left(
 \begin{matrix}
  {\rm e}^{{\rm i}\varphi_2}\:{\rm cos}\:\beta\\
  {\rm e}^{{\rm i}\varphi_2'}\:{\rm sin}\:\beta
 \end{matrix}
 \right)
 \:\:\equiv\:\:
 v_2
 \left(
 \begin{matrix}
  {\rm e}^{{\rm i}\varphi_2}\:c_\beta\\
  {\rm e}^{{\rm i}\varphi_2'}\:s_\beta
 \end{matrix}
 \right),
\end{eqnarray}
\end{subequations}
where $v_1,v_2$ are positive numbers and $\varphi_1,\varphi_1',\varphi_2,
\varphi_2'$ denote the phases of the VEVs. For convenience, we will
work with the relative phases $\varphi\equiv\varphi_1'-\varphi_1$ and
$\psi\equiv\varphi_2'-\varphi_2$. For all potentials $V_0(\Phi)$ which
appear in Eq.~(\ref{eq:V}) we take the quadratic
couplings $A_0$, defined in Eq.~(\ref{eq:V_0}), to be negative. Furthermore, we
assume for all potentials $V_3(\Phi,\Omega)$ in Eq.~(\ref{eq:V})
the quartic couplings $A_3$, defined in Eq.~(\ref{eq:V_3}), to be positive
and sufficiently large compared with the remaining quartic couplings
in Eq.~(\ref{eq:V}). This ensures non-vanishing VEVs $v_1,v_2\neq 0$ and vacuum
stability. Then, the possible physical vacua can be analyzed using the
parameterization of Eqs.~(\ref{eq:parameterization}) where  $v_1$ and $v_2$ are
kept fixed, {\it i.e.}, one can treat each of the fields $\Phi$ and $\Omega$ as
a non-linear $\sigma$-model field and restrict the analysis to the
$(\alpha,\beta,\varphi,\psi)$-parameter-subspace.

In Eqs.~(\ref{eq:V_0}) and (\ref{eq:V_3}) we observe that the potentials
$V_0(\Phi)$ and $V_3(\Phi,\Omega)$ exhibit an accidental
$U(1)_{\rm acc}^4$-symmetry where
$U(1)_{\rm acc}^4 \equiv U(1)_\alpha\times U(1)_\beta\times U(1)_\varphi\times
U(1)_\psi$. As a consequence, the potentials $V_0(\Phi)$ and
$V_3(\Phi,\Omega)$ which appear in
Eq.~(\ref{eq:V}) have no influence on the vacuum alignment\footnote{More
generally, one can view the parameters
$\alpha,\beta,\varphi,$ and $\psi$ as the VEVs of scalar fields
$\tilde{\alpha}(x),\tilde{\beta}(x),\tilde{\varphi}(x),$ and
$\tilde{\psi}(x)$, respectively. The scalar field $\tilde{\alpha}(x)$,
{\it e.g.}, is then the coordinate of the manifold of cosets
$U(1)^4_{\rm acc}/U(1)_{\alpha}$ (which is trivial here)
at each point of space-time. An
alignment of the VEVs of $\Phi$ and $\Omega$ with respect to $\alpha$ happens
when in the lowest energy state $\tilde{\alpha}(x)$ provides only a
{\it non-linear} realization of the group $U(1)^4_{\rm acc}$
(corresponding statements apply to the fields
$\tilde{\beta}(x),\tilde{\varphi}(x),$ and $\tilde{\psi}(x)$).}
of the SM singlet scalars. Therefore, we can without loss of generality discard them for
the rest of our discussion.
 
Assume first that the fields $\Phi$ and $\Omega$ both transform according
to the same irrep $D_i$ $(i=1,2,3,4)$. For
the most general potential involving
only these fields we will denote by $V_\Delta(\Phi,\Omega)$ the part which
breaks the $U(1)_{\rm acc}^4$ symmetry. From Eq.~(\ref{eq:V_1}) we find that
$V_\Delta(\Phi,\Omega)$ can be organized as a sum $V_\Delta(\Phi,\Omega)=V_A(\Phi,\Omega)
+V_B(\Phi,\Omega)$ of two potentials which explicitly read
\begin{eqnarray}\label{eq:V_AV_B}
V_A(\Phi,\Omega)&\equiv&
 d_1|\phi_a^\dagger\phi_b|^2+d_2|\omega_a^\dagger\omega_b|^2+
 d_3(|\phi_a|^2-|\phi_b|^2)(|\omega_a|^2-|\omega_b|^2),\nonumber\\
&&\nonumber\\
V_B(\Phi,\Omega)&\equiv& d_4\left[(\phi_a^\dagger\phi_b)^2+
(\phi_b^\dagger\phi_a)^2\right]
+d_5\left[(\omega_a^\dagger\omega_b)^2+(\omega_b^\dagger\omega_a)^2\right]\nonumber\\
&&+d_6(\phi_a^\dagger\phi_b+\phi_b^\dagger\phi_a)
         (\omega_a^\dagger\omega_b+\omega_b^\dagger\omega_a)\nonumber\\
&&+d_7(\phi_a^\dagger\phi_b-\phi_b^\dagger\phi_a)
         (\omega_a^\dagger\omega_b-\omega_b^\dagger\omega_a),
\end{eqnarray}
where the coefficients $d_1,\ldots,d_7$ are some real-valued numbers.
The potential $V_A(\Phi,\Omega)$ depends only on the angles
$\alpha$ and $\beta$ whereas $V_B(\Phi,\Omega)$ is, in addition, also a function
of $\varphi$ and $\psi$. If the fields satisfy $\Phi\in\{Q^1_1,Q^2_1\}$ and 
$\Omega\in\{Q^1_2,Q^2_2,Q^2_3,Q^2_4\}$ the analogous $U(1)^4_{\rm
acc}$-breaking term $V_B(\Phi,\Omega)$ must have $d_6=d_7=0$. In this case, the
relative phases $\varphi$ and $\psi$ are not correlated in the lowest energy
state. In Eq.~(\ref{eq:V}) we will assume for each
of the different $U(1)^4_{\rm acc}$-symmetry breaking parts $V_A(\Phi,\Omega)$
that $d_1,d_2<0$ and that the condition
\begin{subequations}\label{eq:constraints}
\begin{equation}\label{eq:constraint1}
 d_1d_2>4d_3^2,
\end{equation}
is satisfied. Additionally, we assume that for all possible terms
$V_B(\Phi,\Omega)$ in Eq.~(\ref{eq:V}) the coefficients $d_4$ and $d_5$ are
negative and that they also obey the constraint
\begin{equation}\label{eq:constraint2}
 (-2d_4v_1^4+|d_6|v_1^2v_2^2)
 (-2d_5v_2^4+|d_6|v_1^2v_2^2)>d_7^2v_1^4v_2^4.
\end{equation}
\end{subequations}
As shown in Appendix \ref{app:minimization}, the conditions formulated in
Eqs.~(\ref{eq:constraints}) enforce the nonzero 
VEVs of the component fields to satisfy the relations
\begin{subequations}\label{eq:VEVs}
\begin{eqnarray}
\langle\xi_{ia}\rangle & = & \pm\:\langle \xi_{ib}\rangle\qquad
(i=1,2),\label{eq:xiVEVs}\\
\langle q^1_{ja}\rangle&=& \pm\:\langle q^1_{jb}\rangle\qquad (j=1,2),\\
\langle q^2_{ka}\rangle&=&\pm\:\langle q^2_{kb}\rangle\qquad (k=1,2,3,4),
\end{eqnarray}
\end{subequations}
{\it i.e.}, within each of the scalar $\mathscr{G}$-doublets the VEVs of the
component fields are relatively real and exactly degenerate
(up to a possible relative sign).\footnote{We consider here only
the tree-level approximation.}
Note that all terms in the potentials $V_B(\Phi,\Omega)$ which are multiplied by the 
coefficient $d_7$ must vanish in the lowest energy state and, hence, cannot
contribute to the minimization of the scalar potential. For the potential $V$
we furthermore choose in both of the terms $V_B(Q^1_1,Q^2_1)$ and
$V_B(\xi_1,\xi_2)$ the corresponding coefficients $d_6$ to be positive. In contrast to this, the
non-vanishing coefficients $d_6$ in all of the remaining terms $V_B(\Phi,\Omega)$ of $V$
are all assumed to be negative. Then, the absolute minimum of the multi scalar
potential is in addition to Eqs.~(\ref{eq:VEVs}) furthermore characterized by the
relations
\begin{subequations}\label{eq:vacuumorientation}
\begin{eqnarray}
 &&
 \frac{\langle\xi_{1a}\rangle}{\langle\xi_{1b}\rangle}=
 -\frac{\langle\xi_{2a}\rangle}{\langle\xi_{2b}\rangle},\label{eq:xiorientation}\\
 &&
 \frac{\langle q^1_{1a}\rangle}{\langle q^1_{1b}\rangle}=-
 \frac{\langle q^2_{1a}\rangle}{\langle q^2_{1b}\rangle},\\
 &&
 \frac{\langle q^1_{2a}\rangle}{\langle q^1_{2b}\rangle}=
 \frac{\langle q^2_{2a}\rangle}{\langle q^2_{2b}\rangle}=
 \frac{\langle q^2_{3a}\rangle}{\langle q^2_{3b}\rangle}=
 \frac{\langle q^2_{4a}\rangle}{\langle q^2_{4b}\rangle},
\end{eqnarray}
\end{subequations}
{\it i.e.}, the relative orientation of the VEVs of the component fields within
a specific doublet is equal for all doublets transforming under the irrep
$D_4$ and opposite for the pairs of doublets transforming under $D_1$ or
$D_3$.\footnote{The
terms associated with the coefficients $d_6$ actually represent a
spin-spin-interaction in a version of the Ising-model, known from
ferromagnetism. The topology here is of course unfamiliar since all ``spins'' couple
with equal strength.}

We suppose that the VEVs of the fields $\xi_0,\xi_1,$ and $\xi_2$, which
are responsible for the generation of the neutrino masses, are all of the order
of the electroweak scale
\begin{equation}\label{eq:hierarchy}
 |\langle\xi_{0}\rangle|\simeq|\langle\xi_{1a}\rangle|\simeq
 |\langle\xi_{2a}\rangle|\simeq 10^2\:{\rm GeV}.
\end{equation}
In contrast to this, we assume that the link fields of the deconstructed
extra-dimensional gauge symmetries $G^1$ and $G^2$ all aquire VEVs of the same order of magnitude 
at some high mass scale somewhat below $M_f$ and thereby give rise to a small expansion
parameter
\begin{equation}\label{eq:epsilonl}
\lambda\simeq\frac{|\langle q^1_{ja}\rangle|}{M_f}
\simeq\frac{|\langle q^2_{ka}\rangle|}{M_f}\simeq 0.22,
\end{equation}
where $j=1,2,$ and $k=1,2,3,4$. Small hierarchies of this type can
emerge from large hierarchies in supersymmetric theories when the scalar fields
aquire their VEVs along a D-flat direction \cite{witt81}. Note in
Eq.~(\ref{eq:epsilonl}) that $\lambda$ is given by 
the Wolfenstein parameter \cite{wolf83} which approximately describes the mass
ratios and CKM mixing angles in the down-quark sector \cite{ros001} as well as
the mass ratios in the charged lepton sector \cite{groo00}.\footnote{An Ansatz where the
Wolfenstein parameter is also used to describe neutrino mixing and leptogenesis
has recently been presented in Ref.~\cite{Xing:2002kz}.} The mass and mixing
parameters of the charged leptons are determined in the next section.

\section{The charged lepton mass matrix}\label{sec:chargedleptons}
Consider the Yukawa interactions of the charged leptons
\begin{equation}
\mathscr{L}^\ell_Y=\overline{L_\alpha}H\mathscr{O}^\ell_{\alpha\beta}E_\beta+
\rm{h.c.},
\end{equation}
where $\mathscr{O}^\ell_{\alpha\beta}$ denotes an effective Yukawa operator
and $\alpha,\beta=e,\mu,\tau$.
The $G_L\times G_R$ charge structure of the charged lepton-antilepton pairs is
shown in Table \ref{tab:chargedleptons}.
\begin{table}
\begin{center}
\begin{tabular}{c|c|ccc}
& & $E_e$ & $E_\mu$ & $E_\tau$\\ \hline
	      & $G_L\times G_R$ & $(-2,2)$ & $(0,1)$ & $(0,1)$\\ \hline
$\overline{L_e}$   & $(-2,2)$ & $(-4,4)$ & $(-2,3)$ & $(-2,3)$ \\
$\overline{L_\mu}$ & $(-1,0)$ & $(-3,2)$ & $(-1,1)$ & $(-1,1)$ \\
$\overline{L_\tau}$& $(-1,0)$ & $(-3,2)$ & $(-1,1)$ & $(-1,1)$ \\
\end{tabular}
\end{center}
\caption{$G_L\times G_R$ charge structure of the charged 
lepton-antilepton pairs.}
\label{tab:chargedleptons}
\end{table}
One of the possible lowest-dimensional contributions to the operator
$\mathscr{O}^\ell_{ee}$ is given by
\begin{equation}\label{eq:Oee}
 \mathscr{O}^\ell_{ee}\simeq(Q^1_1Q^1_2)^4/(M_f)^8,
\end{equation}
since it yields an invariant under both $G_L\times G_R$ and $\mathscr{G}$.
Furthermore, invariance under application of the operator $\mathcal{D}(a_4)$ requires
$\mathscr{O}^\ell_{e\mu}=\mathscr{O}^\ell_{\mu e}=0$. Since the transformation  $\mathcal{D}(a_2)$ permutes
$\mathscr{O}^\ell_{e\mu}\leftrightarrow\pm\mathscr{O}^\ell_{e\tau}$
and $\mathscr{O}^\ell_{\mu e}\leftrightarrow\pm\mathscr{O}^\ell_{\tau e}$
(a sign flip is possible due to $D_3(a_2)=D(C_b)$) it follows that 
$\mathscr{O}^\ell_{e\tau}=\mathscr{O}^\ell_{\tau e}=0$, {\it i.e.}, in the first
row and column of the charged lepton mass matrix only the $(1,1)$-element is
non-vanishing. From $U(1)$ gauge-invariance we have seen in Sec.~\ref{sec:U(1)charges}
that the lowest-dimensional mass terms in the $\mu$-$\tau$-subsector of the
charged leptons are given by the dimension-six and dimension-eight operators
shown in Fig.~\ref{fig:muontaumasses}. In theory space, the effective Yukawa
couplings of the dimension-six and dimension-eight operators are identified with
the Wilson loops around the plaquettes associated with the deconstructed 
extra-dimensional gauge symmetries $G^1$ and $G^2$ (see Figs.~\ref{fig:contour1} and
\ref{fig:contour2}). The generalized
Wigner-Eckart theorem implies that each of these Wilson loops corresponds in
$\mathscr{G}$-space to a set of irreducible Yukawa tensor operators spanned by
the irreps $\mathbf{2}_\ell$ and $\mathbf{2}_E$. These operators can be quickly
determined by first noting that under $\mathcal{D}(a_4)$ the effective couplings
$\mathscr{O}^\ell_{\mu\tau}$ and
$\mathscr{O}^\ell_{\tau\mu}$ undergo a sign flip whereas all link fields of
$\Pi^2_{i=1}G^1_i$ and $\Pi^4_{i=1}G^2_i$ transform trivially. As a result, we have 
$\mathscr{O}^\ell_{\mu\tau}=\mathscr{O}^\ell_{\tau\mu}=0$, {\it i.e.},
the charged lepton mass matrix is diagonal.
Then, testing for invariance under the action of
$a_1,a_2,a_3\in\mathscr{G}$ gives that each set
of irreducible Yukawa tensor operators transforms according to a representation
$D_5$ of $\mathscr{G}$ which is in matrix-form defined by the generators
\begin{equation}\label{eq:D5}
 D_5(a_1)=
 \left(
 \begin{matrix}
 0 & 1\\
 1 & 0
 \end{matrix}
 \right),\quad
  D_5(a_2)=
 \left(
 \begin{matrix}
 -1 & 0\\
 0 & 1
 \end{matrix}
 \right),\quad
 D_5(a_3)=D_5(a_4)=\mathbbm{1}_2.
\end{equation}
To leading order, the generators in Eq.~(\ref{eq:D5})
act on five independent doublets $(\psi_a^{(i)},\:\psi_b^{(i)})^T$ of product functions
$\psi_a^{(i)}$ and $\psi_b^{(i)}$, where $i=0,1,\ldots,4$, which form the basis
of five distinct carrier spaces $V^{(i)}$. Here, the $\mathscr{G}$-doublets
$(\psi_a^{(i)},\:\psi_b^{(i)})^T$ correspond for $i=0$ and $i\geq 1$ to the
Wilson loops around the plaquettes associated with $G^1$ and
$G^2$, respectively. The basis functions for the different
carrier spaces of $D_5$ are given in table \ref{tab:basisfunctions}.
\begin{table}
\begin{center}
\begin{tabular}{|c|cc|}
\hline
$V^{(i)}$ & $\psi_a^{(i)}$ & $\psi_b^{(i)}$ \\ \hline\hline
$V^{(0)}$   & $q^1_{1a}q^1_{2a}$ &
                     $q^1_{1b}q^1_{2b}$\\
$V^{(1)}$   & $q^2_{1a}q^2_{2a}q^2_{3a}q^2_{4a}$ &
                     $q^2_{1b}q^2_{2b}q^2_{3b}q^2_{4b}$\\
$V^{(2)}$ & $q^2_{1a}q^2_{2a}q^2_{3b}q^2_{4b}$ &
                     $q^2_{1b}q^2_{2b}q^2_{3a}q^2_{4a}$ \\
$V^{(3)}$& $q^2_{1a}q^2_{2b}q^2_{3a}q^2_{4b}$ &
                     $q^2_{1b}q^2_{2a}q^2_{3b}q^2_{4a}$ \\
$V^{(4)}$& $q^2_{1a}q^2_{2b}q^2_{3b}q^2_{4a}$ &  
                   $q^2_{1b}q^2_{2a}q^2_{3a}q^2_{4b}$ \\\hline
\end{tabular}
\end{center}
\caption{The basis functions $\psi_a^{(i)}$ and $\psi_b^{(i)}$
$(i=0,1,\ldots,4)$ of the
leading-order set of irreducible Yukawa tensor operators
$(\mathscr{O}^\ell_{kl})$, where $k,l=\mu,\tau$, in terms of the component
fields of $Q^1_{i}=(q^1_{ia},\:q^1_{ib})^T$ $(i=1,2)$ and
$Q^2_{i}=(q^2_{ia},\:q^2_{ib})^T$ $(i=1,2,3,4)$.}
\label{tab:basisfunctions}
\end{table}
The allowed types of basis functions in table \ref{tab:basisfunctions} follow
from invariance under application of the transformations
$\mathcal{D}(a_3)$ and $\mathcal{D}(a_1)$. First, application of $\mathcal{D}(a_3)$ yields that
in each basis function the number of the $a$-components is even. Second, for
given $i=0,1,\ldots,4$
invariance under the transformation $\mathcal{D}(a_1)$ requires the index structure
of the basis functions $\psi^{(i)}_a$ and $\psi^{(i)}_b$ to be of such
a form that $\psi^{(i)}_a$ and $\psi^{(i)}_b$ get interchanged when
the indices $a$ and $b$ are permuted.
Then, expressed in terms of the sets of irreducible Yukawa tensor operators, the
effective Yukawa coupling matrix in the $\mu$-$\tau$-subsector of the charged
leptons reads
\begin{subequations}\label{eq:tensoroperator}
\begin{equation}\label{eq:tensoroperator1}
 (\mathscr{O}^\ell_{kl})=
 \sum_{i=0}^4 Y^{(i)}_{\rm eff}\left[
 \left(
 \begin{matrix}
  1 & 0\\
  0 & 1
 \end{matrix}
 \right)\psi^{(i)}_a
 +
 \left(
 \begin{matrix}
  -1 & 0\\
  0 & 1
 \end{matrix}
 \right)\psi^{(i)}_b
 \right],
\end{equation}
where $k,l=\mu,\tau$ and $Y^{(i)}_{\rm eff}$ for $i=0,1,\ldots,4$ denotes
the effective Yukawa couplings
\begin{equation}
 Y^{(i)}_{\rm eff}
 =\left\{
 \begin{matrix}
  Y^{(0)}/(M_f)^2 & & (i=0)\qquad\:\:\:\\
  &&\\
  Y^{(i)}/(M_f)^4 & & (i=1,2,3,4)
 \end{matrix}
 \right\},
\end{equation}
\end{subequations}
where $Y^{(i)}$ $(i=0,1,\ldots,4)$ are order unity Yukawa couplings.
The $2\times 2$ diagonal matrices in Eq.~(\ref{eq:tensoroperator1}) summarizing
symmetry-related geometric factors, are the Clebsch-Gordan coefficients of the
effective Yukawa coupling matrix $(\mathscr{O}^\ell_{kl})$. Furthermore, the
effective Yukawa couplings $Y^{(i)}_{\rm eff}$, characterized by the outer
multiplicity label $i$, are the reduced matrix elements
of the Clebsch-Gordan coefficients and parameterize further information
about the physics at the fundamental scale $M_f$.
From Eqs.~(\ref{eq:VEVs}) we find that after SSB the vacuum alignment mechanism
of Sec.~\ref{sec:vacuumalignment} ensures that the VEVs of the basis functions
in table \ref{tab:basisfunctions} are - up to a possible relative sign - pairwise
exactly degenerate
\begin{subequations}\label{eq:psiVEVs}
\begin{equation}
 \langle\psi^{(i)}_a\rangle=\pm \langle\psi^{(i)}_b\rangle,
\end{equation}
where $i=0,1,\ldots,4$. In addition, Eqs.~(\ref{eq:vacuumorientation}) relate
the orientations of the VEVs by
\begin{equation}
\langle\psi^{(0)}_a\rangle/\langle \psi^{(0)}_b\rangle=-
\langle\psi^{(i)}_a\rangle/\langle\psi^{(i)}_b\rangle,
\end{equation}
\end{subequations}
where $i=1,2,3,4$. Substituting Eqs.~(\ref{eq:psiVEVs}) into
Eqs.~(\ref{eq:tensoroperator}) we observe that after SSB the set of irreducible
Yukawa tensor operators $(\mathscr{O}^\ell_{kl})$ can take one of the following 
two forms
\begin{equation}\label{eq:mutaugeneration}
(\mathscr{O}^\ell_{kl})\rightarrow
 Y^{(0)}\lambda^2\left[
 \left(
 \begin{matrix}
  1 & 0\\
  0 & 1
 \end{matrix}
 \right)
 \pm
 \left(
 \begin{matrix}
  -1 & 0\\
  0 & 1
 \end{matrix}
 \right)
 \right]
 +
 \sum_{i=1}^4
 Y^{(i)}\lambda^4\left[
 \left(
 \begin{matrix}
  1 & 0\\
  0 & 1
 \end{matrix}
 \right)
 \mp
 \left(
 \begin{matrix}
  -1 & 0\\
  0 & 1
 \end{matrix}
 \right)
 \right],
\end{equation}
where $k,l=\mu,\tau$ and the expansion parameter $\lambda\simeq 0.22$ of
Eq.~(\ref{eq:epsilonl}) has been used. In Eq.~(\ref{eq:mutaugeneration}) it is
important to note that the Clebsch-Gordan coefficients are added or subtracted
depending on their outer multiplicity label: if the Clebsch-Gordan
coefficients are added (subtracted) for $i=0$ then they are necessarily
subtracted (added) for $i=1,2,3,4$. As a result, Eq.~(\ref{eq:mutaugeneration})
shows that the vacuum alignment mechanism generates a hierarchical pattern in
the $\mu$-$\tau$-subsector of the charged leptons via a cancellation of some
of the Clebsch-Gordan coefficients in the lowest energy state. For definiteness,
let us choose in Eq.~(\ref{eq:mutaugeneration}) the solution with the signs
``$+$'' for $i=0$ and
``$-$'' for $i=1,2,3,4$.
Taking everything into account, the full leading order charged
lepton mass matrix $\mathcal{M}_\ell$ emerging after SSB is given by
\begin{equation}\label{eq:M_l}
\mathcal{M}_\ell\simeq m_\tau
\left(
\begin{matrix}
 \lambda^6 & 0 & 0\\
 0 & \lambda^2 & 0\\
 0 & 0 & 1
\end{matrix}
\right),
\end{equation}
where $m_\tau$ is the tau mass and
only the orders of magnitude of the matrix elements have been indicated. The
masses and the mixing of the charged leptons can be calculated by
diagonalizing $\mathcal{M}_\ell\mathcal{M}_\ell^\dagger$. Denoting the electron
and muon masses by $m_e$ and $m_\mu$, respectively, the mass spectrum described
by $M_\ell$ is found to be
\begin{equation}\label{eq:chargedleptonmasses}
 \frac{m_e}{m_\tau}\simeq \lambda^6,\quad \frac{m_\mu}{m_\tau}\simeq
 \lambda ^2,
\end{equation}
which approximately fits the experimentally observed values \cite{groo00}.
As for the mixing angles of the charged
leptons practically vanish, the experimentally observed large leptonic mixing
must stem from the neutrino sector. The neutrino mass and mixing parameters
will be determined in the next section.

\section{The neutrino mass matrix}\label{sec:neutrinos}
\subsection{Aliphatic models}\label{sec:orbifold}
So far, we have examined the connection between dynamically generated extra
dimensions compactified on $\mathcal{S}^1$ and the hierarchical Yukawa coupling
matrix of the charged leptons. In such a geometric approach to small Yukawa
couplings it is generally interesting to relate the leptonic mass and mixing parameters to different
topologies in theory space. For this purpose we will now, following
Sec.~\ref{sec:aliphatic}, consider for each of the gauge groups
$G^3=\Pi^{N_3}_{i=1}G^3_i$ and $G^4=\Pi^{N_4}_{i=1}G^4_i$ the
aliphatic model for fermions in the latticized $\mathcal{S}^1/Z_2$ orbifold
extra dimensions. Since these fermions are SM singlets, we can identify them
with ``right-handed'' ({\it i.e.}, $SU(2)_L$ singlet) neutrinos. We suppose that
the fundamental Froggatt-Nielsen states of the charged lepton sector transform
only trivially under $G^3$ and $G^4$ (see Sec.~\ref{sec:U(1)charges}).
Then, the orbifold extra dimensions are (at tree-level) completely decoupled
from the charged leptons and can only be experienced by the neutrinos.

\begin{figure}
\begin{center}
\includegraphics*[bb = 93 583 481 691]{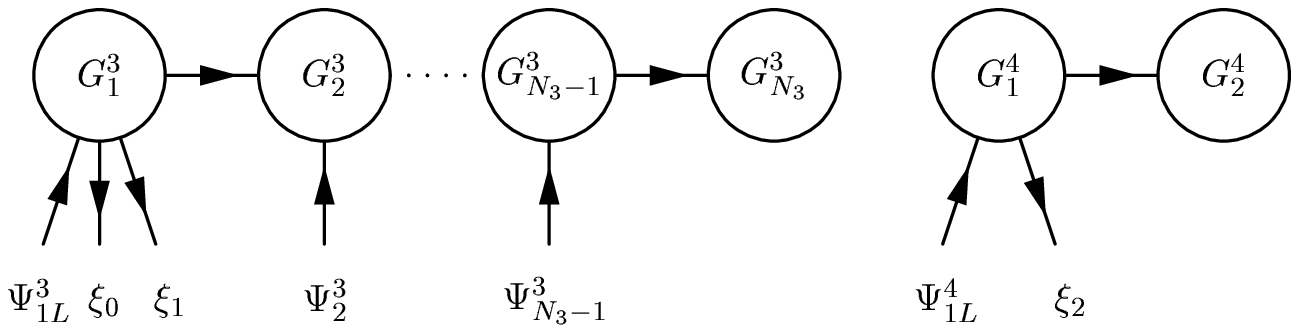}
\end{center}
\vspace{-3mm}
\caption{\small{Moose diagram for the location of scalars and chiral fermions
   in the orbifold extra dimensions associated with the gauge groups
    $G^3$ and $G^4$.}}
  \label{fig:sites}
\end{figure}

At this stage, we allow the number of sites $N_3$ of the gauge group $G^3$ to be
large but leave it yet unspecified\footnote{In Sec.~\ref{sec:neutrinoparameters}
we will show that the number of sites $N_3$ parameterizes the solar neutrino
mass squared difference $\Delta m_\odot^2$.}.
For the gauge group $G^4$, however, we assume a very coarse latticization where
the aliphatic ``chain'' of $G^4$ consists only of the $N_4=2$ sites $G^4_1$ and
$G^4_2$. By assigning $\xi_0$ and $\xi_1$ the $G^3_1$
charge $+1$ these fields are put on the 1st site associated with $G^3$.
Additionally, we assign $\xi_2$ the $G^4_1$ charge $+1$ which locates the
field at the site representing $G^4_1$. The corresponding moose diagrams are
shown in Fig.~\ref{fig:sites}.
Since the charged leptons cannot experience these topologies, the
site-fields $\xi_0,\xi_1,$ and $\xi_2$ are now only relevant for the generation
of the neutrino mass matrix and we can discard them in the discussion of
the charged lepton mass matrix.

\subsection{The one-generation-case}\label{sec:onegeneration}
Let us first restrict to the case of one generation by considering
the coupling of the $\mathscr{G}$-singlet neutrino $\nu_e$ to the fermionic
site variables of the orbifold extra dimensions. From table 
\ref{tab:GLGRcharges} we conclude that $G_L\times G_R$ gauge-invariance requires
the relevant effective Yukawa interaction of the electron lepton doublet
$\mathbf{1}_\ell$ with the bulk-fermions to be of the
type
\begin{eqnarray}\label{eq:L_2}
 {\mathscr{L}}_2& = &
 Y_1\overline{\mathbf{1}^c_\ell}\tilde{H}^\ast\mathbf{1}_{FL}+Y_2
 \overline{\mathbf{1}}_{FR}\xi_0\Psi^3_{1L}+
 M_{1F}\overline{\mathbf{1}}_{FR}\mathbf{1}_{FL}+{\rm h.c.},
\end{eqnarray}
where $Y_1$ and $Y_2$ are order unity coefficients, $\tilde{H}$ is the
charge-conjugated Higgs-field $\tilde{H}={\rm i}\sigma^2 H^\ast$, and
$M_{1F}\simeq 10^{15}\:{\rm GeV}$ denotes the seesaw scale.
Suppressing the index $j=3$, the Lagrangian
for the bulk and brane fields is given by $\mathscr{L}_1+\mathscr{L}_2$, where
$\mathscr{L}_1$ has been defined in Eq.~(\ref{eq:L_1}). As a result,
the $N_3\times N_3$ mass matrix which emerges from 
${\mathscr{L}}_1+{\mathscr{L}}_2$ after SSB is given by
\begin{equation}\label{eq:completesystem}
\begin{matrix}
&
\begin{matrix}
 \mathbf{1}_{FL} & \Psi_{1L} & \Psi_{2L} & \Psi_{3L} & \cdots 
\end{matrix}\vspace{1mm}
\\
  \begin{matrix}
   \overline{\mathbf{1}^c_\ell}\\
   \overline{\mathbf{1}}_{FR}\\
   \overline{\Psi}_{2R}\\
   \overline{\Psi}_{3R}\\
   \vdots
  \end{matrix}
  &
  \left(\hspace{3mm}
  \begin{matrix}
   h_1 & 0 & 0 & 0 & \cdots\\
   M_{1F} & h_2 & 0 & 0 & \cdots \\
   0 & M_f & -M_f & 0& \cdots \\
   0 & 0 & M_f & -M_f & \cdots \\
   \vdots & \vdots & \vdots & \vdots & \ddots
  \end{matrix}\hspace{3mm}
  \right)
 \end{matrix},
\end{equation}
where we have introduced the VEVs $h_1\equiv Y_1 \langle \tilde{H}^\ast\rangle$,
$h_2\equiv Y_2\langle \xi_0\rangle$, and use $h_1\simeq h_2\simeq 10^2\:{\rm
GeV}$, {\it i.e.}, the associated mass terms are generated at the electroweak
scale. Note that the mass matrix in Eq.~(\ref{eq:completesystem})
has been displayed in a basis where the active neutrinos have right-handed
chirality. Integrating out the heavy $\mathscr{G}$-singlet $\mathbf{1}_F$, yields
the effective $(N_3-1)\times (N_3-1)$ mass matrix
\begin{equation}\label{eq:M_2}
{\mathcal{M}}_2\equiv
\left(
\begin{matrix}
   m_\nu & 0 & 0 & \cdots \\
   M_f & -M_f & 0& \cdots \\
   0 & M_f & -M_f & \cdots \\
   \vdots & \vdots & \vdots & \ddots \\
  \end{matrix}
 \right),
\end{equation}
where $m_\nu=h_1h_2/M_{1F}\simeq 10^{-3}\:{\rm eV}$ denotes the absolute neutrino
mass scale. The mass $m_\nu$ lifts the zero eigenvalue in the
KK mass spectrum of the left-handed fields to a small but
non-vanishing value which can be determined by diagonalizing the
$(N_3-1)\times (N_3-1)$ matrix
\begin{equation}\label{eq:M_2^2}
 {\mathcal{M}}_2{\mathcal{M}}_2^\dagger=
 \left(
 \begin{matrix}
  m_\nu^2 & m_\nu M_f & 0&\cdots\\
  m_\nu M_f & 2M_f^2 & -M_f^2&\cdots\\
  0 & -M_f^2 & 2M_f^2&\cdots\\
  \vdots & \vdots &\vdots & \ddots
 \end{matrix}
 \right).
\end{equation} 
From Eq.~(\ref{eq:M_2^2}) it is readily seen that the mixing of
$\mathbf{1}^c_\ell$ with the right-handed fields $\Psi_{nR}$
is described by angles $\simeq m_\nu/M_f$ which vanish in the limit
$M_f\rightarrow \infty$. For $m_\nu\ll M_f$ the non-zero heavy masses are
approximately given by the KK spectrum in Eq.~(\ref{eq:M_n,L}) where
$n=1,\ldots,N_3-2$. The lightest eigenvalue $(m_0)^2$ of
${\mathcal{M}}_2{\mathcal{M}}_2^\dagger$
can be determined by integrating out the invertible
heavy $(N_3-2)\times (N_3-2)$ submatrix in the down-right corner
of ${\mathcal{M}}_2{\mathcal{M}}_2^\dagger$ in Eq.~(\ref{eq:M_2^2}).
Taking into account that the $(1,1)$-element of the
inverse of this submatrix is equal to $\frac{N_3-2}{N_3-1}M_f^{-2}$, we obtain in the
limit $m_\nu\ll M_f$ for the lightest mass eigenvalue
\begin{equation}\label{eq:finestructure}
 m_0=m_\nu/\sqrt{N_3-1}= m_\nu/\sqrt{M_f R} ,
\end{equation}
where the second equation matches onto the continuum limit. In the classical 5D
continuum theory the mass (or volume) suppression factor $\sim\sqrt{M_f R}$ emerges from the
normalization of the wave-function of the right-handed neutrino propagating in
the bulk \cite{dien99}. In our deconstruction setup,
Eq.~(\ref{eq:finestructure}) states that for a given number of sites there is a
non-decoupling of the deconstruction physics from the low-energy theory.
Since, as already stated after Eq.~(\ref{eq:M_2^2}), the mixing of the
active SM lepton $\mathbf{1}^c_\ell$ with the right-handed fields $\Psi_{nR}$ vanishes in the
limit $M_f\rightarrow \infty$, we can repeat the above construction for
additional aliphatic models which, after integrating out the corresponding KK
towers, give rise to further light Dirac neutrino masses. In this way, one can
build up a Dirac neutrino mass matrix, which reflects the properties
of the dynamically generated extra dimensions.

\subsection{Inclusion of the 2nd and 3rd generations}
The full neutrino mass matrix emerges by inclusion of the $SU(2)$ lepton
doublets of the 2nd and 3rd generation, which are combined into the
$\mathscr{G}$-doublet
$\mathbf{2}_\ell$. Actually, gauge-invariance under $G_L\times G_R$ allows for
the fields $\xi_2$ and $\xi_3$ Yukawa interactions with $\mathbf{2}_\ell$ which read
\begin{eqnarray}
{\mathscr{L}}_3& = &
 Y_3\overline{\mathbf{2}^c_\ell}\tilde{H}^\ast\mathbf{2}_{FL}+Y_4
 \overline{\mathbf{2}}_{FR}\xi_1\Psi^3_{1L}
 +Y_5\overline{\mathbf{2}}_{FR}\xi_2\Psi^4_{1L}
 +M_{2F}\overline{\mathbf{2}}_{FR}\mathbf{2}_{FL}+{\rm h.c.},
\end{eqnarray}
where $Y_3,Y_4,$ and $Y_5$ denote order unity Yukawa couplings,
$\langle\xi_1\rangle\simeq\langle\xi_2\rangle\simeq 10^2\:{\rm GeV}$, and 
$M_{2F}\simeq 10^{15}\:{\rm GeV}$.
In the construction of Sec.~\ref{sec:aliphatic} we associate with
the Weyl spinor $\Psi^4_{1L}$ a two-site model
for $G^4$ where $\Psi^4_{1L}$ resides on the site
corresponding to $G^4_1$ (see
Fig.\ref{fig:sites}). In complete analogy to the calculation of the light mass
$m_0$ in Sec.~\ref{sec:onegeneration}, one can in the
combined system $(\sum_{i=3,4}\mathscr{L}_1)+{\mathscr{L}}_2+{\mathscr{L}}_3$,
where $\sum_{i=3,4}\mathscr{L}_1$ summarizes the aliphatic chains of
$G^3$ and $G^4$, integrate out the heavy
vectorlike degrees of freedom. As a consequence, in the basis where the VEVs of
$\xi_1$ and $\xi_2$ are described by Eqs.~(\ref{eq:xiVEVs}) and
(\ref{eq:xiorientation}) the resulting $3\times 2$ light Dirac neutrino mass
matrix can be written as
\begin{equation}\label{eq:M_D}
 {\mathcal{M}}_D= m_\nu
 \left(
 \begin{matrix}
  \rho\epsilon & 0\\
  \epsilon & 1\\
  -\epsilon &1
 \end{matrix}
 \right),
\end{equation}
where the neutrino expansion parameter $\epsilon\simeq 1/\sqrt{N_3-1}$ and the
order unity Yukawa coupling $\rho=\mathcal{O}(1)$ are both real quantities and
$m_\nu$ denotes the absolute neutrino mass scale. In Eq.~(\ref{eq:M_D}) all
phases have been absorbed into the right-handed neutrino and charged lepton
sectors\footnote{This is possible since the charged lepton mass matrix
$\mathcal{M}_\ell$ in Eq.~(\ref{eq:M_l}) is of diagonal form.} implying
the practical absence of $\mathcal{CP}$ violation in neutrino oscillations. The
Dirac neutrino mass matrix $\mathcal{M}_D$ in Eq.~(\ref{eq:M_D}) has the
important property that within each column the flavor symmetry $\mathscr{G}$
enforces the 2nd and 3rd elements to be relatively real and exactly degenerate
in their magnitudes. Thus, $\mathcal{M}_D$ describes an exactly maximal
$\nu_\mu$-$\nu_\tau$-mixing. The structure of $\mathcal{M}_D$ is familiar from
models of single right-handed neutrino dominance \cite{king99} which provide an understanding of normal
hierarchical neutrino mass spectra in the context of the MSW LMA solution.

\subsection{Neutrino masses and mixing angles}\label{sec:neutrinoparameters}
The neutrino masses and leptonic mixing angles are determined from
Eq.~(\ref{eq:M_D}) by diagonalizing the matrix
\begin{equation}\label{eq:M_D^2}
 {\mathcal{M}}_D{\mathcal{M}}_D^\dagger=m_\nu^2
 \left(
 \begin{matrix}
  \rho^2\epsilon^2&\rho\epsilon^2&-\rho\epsilon^2\\
  \rho\epsilon^2&1+\epsilon^2&1-\epsilon^2\\
  -\rho\epsilon^2 &1-\epsilon^2&1+\epsilon^2
 \end{matrix}
 \right).
\end{equation}
The matrix in Eq.~(\ref{eq:M_D^2}) is brought on diagonal form by a rotation of
the active neutrino fields in the 2-3-plane through an angle $\theta_{23}=\pi/4$
followed by a rotation in the 1-2-plane through an angle
\begin{equation}
 \theta_{12}={\rm arctan}\left[(2\sqrt{2})^{-1}\left(
 \rho^2-2+\sqrt{(2-\rho)^2+8}\right)\right].
\end{equation}
Hence, the reactor mixing angle $\theta_{13}$ exactly vanishes\footnote{At tree-level,
sub-leading corrections may come from the charged lepton sector.} in agreement
with the CHOOZ reactor neutrino data which sets the upper bound
$|\theta_{13}|\lesssim 9.2^\circ$ \cite{cho98}.
The neutrino masses
exhibit the normal hierarchy
\begin{equation}
 m_1=0,\quad m_2=m_\nu\epsilon\sqrt{2+\rho^2},\quad m_3=\sqrt{2}
 m_\nu,
\end{equation}
which gives for the solar and atmospheric neutrino mass squared differences
\begin{equation}
 \Delta m^2_\odot=m_\nu^2\epsilon^2(2+\rho^2),\quad
 \Delta m^2_{\rm atm}=2m_\nu^2-\Delta m^2_\odot.
\end{equation}
Using the upper bound $\Delta m^2_\odot\lesssim 1.9\times 10^{-4}\:{\rm eV}^2$
\cite{mal002} and the best-fit value $\Delta m_{\rm atm}^2\simeq 2.5\times
10^{-3} \:{\rm eV}^2$ \cite{super0012} we obtain $m_\nu\simeq 0.04\:{\rm eV}$
which is consistent with the value
$M_{F_1}\simeq M_{F_2}\simeq 10^{15}\:{\rm GeV}$ for the seesaw scale.
Without tuning of parameters we have $\rho=1$ and $\epsilon=1/\sqrt{N_3-1}$
which gives for the solar neutrino parameters the values
\begin{equation}\label{eq:solarparameters}
 \Delta m^2_\odot  \simeq  \frac{3}{2}\frac{\Delta m_{\rm atm}^2}{N_3-1},\quad
 \theta_{12}   ={\rm arctan}\:\frac{1}{\sqrt 2} \simeq  35^\circ,
\end{equation}
where we have used in the first equation the hierarchy $\Delta
m_\odot^2\ll\Delta m_{\rm atm}^2$.
At $3\sigma$, the combined solar and KamLAND neutrino data allows 
for $\Delta m_\odot^2$ the two regions\footnote{We adopt here the
nomenclature of Ref.~\cite{fo002}.}
$5.1\times 10^{-5}\:{\rm eV}^2\lesssim \Delta m_\odot^2\lesssim
 9.7\times 10^{-5}\:{\rm eV}^2$ (LMA-I) and
$1.2\times 10^{-4}\:{\rm eV}^2\lesssim
  \Delta m_\odot^2\lesssim 1.9\times 10^{-4}\:{\rm eV}^2$ (LMA-II)
\cite{mal002}.
Matching onto these values requires 
\begin{equation}\label{eq:LMAIandII}
N_3=57\pm 17 \quad \text{LMA-I},\qquad
N_3=27\pm 6 \quad \text{LMA-II},
\end{equation}
where we have set in Eq.~(\ref{eq:solarparameters}) the atmospheric mass squared
difference equal to the best-fit value $\Delta m_{\rm atm}^2 = 2.5\times
10^{-3} \:{\rm eV}^2$. In short, the presently allowed ratios $\Delta
m_\odot^2/\Delta m_{\rm atm}^2$ implied by the LMA-I and LMA-II solutions
already significantly discriminate between the associated radii $N_3/(2\pi gv)$ of the dynamically
generated $\mathcal{S}^1/Z_2$ orbifold. At this level, the neutrino expansion
parameter is
$0.12\lesssim\epsilon\lesssim 0.16$ (LMA-I) or
$0.18\lesssim\epsilon\lesssim 0.23$ (LMA-II), which is comparable with the
Wolfenstein parameter $\lambda\simeq 0.22$ of the charged fermion sector.
For the non-fine-tuned solar mixing angle
$\theta_{12}={\rm arctan}\:1/\sqrt{2}$ in Eq.~(\ref{eq:solarparameters}) we find
from the analysis in Ref.~\cite{bah0022} that a number of
\begin{equation}\label{eq:90}
 55\pm 8\quad\text{LMA-I}\:\:(@\:90\%\:\text{C.L.})
\end{equation}
lattice sites yields the MSW LMA-I solution within the 90\% confidence level
region. In general, for both the LMA-I and the LMA-II solution, the dynamical
generation of the solar mass squared difference $\Delta m_\odot^2$ via
deconstruction in a flat background requires a relatively fine-grained latticization of the
associated $\mathcal{S}^1/Z_2$ orbifold with roughly $10^1-10^2$ lattice sites.

\section{Summary and Conclusions}\label{sec:summaryandconclusions}
In conclusion, we have presented a model for lepton masses and mixing angles
based on a non-Abelian discrete flavor symmetry and $U(1)$ charges in a
deconstructed setup. The model applies the vacuum alignment mechanism of
previous models\cite{ohl0021,ohl0022} in order to predict an exactly maximal
$\nu_\mu$-$\nu_\tau$-mixing as well as the strict hierarchy
$m_\mu \ll m_\tau$ between the muon and the tau mass. The non-Abelian discrete 
symmetry is identified as a split extension of the Klein group $Z_2\times Z_2$
where the lift of every fermion and scalar representation is equivalent with the
dihedral group $\mathscr{D}_4$ of order eight. The charged lepton masses are generated by the
Froggatt-Nielsen mechanism which is given a geometric interpretation in terms of deconstructed or latticized
extra dimensions compactified on the circle $\mathcal{S}^1$. In theory space,
the muon and tau masses correspond to Wilson loops around the plaquettes
associated with the deconstructed extra-dimensional $U(1)$ gauge symmetries. As a
result, the model gives the realistic charged lepton mass ratios
$m_e/m_\tau\simeq \lambda^6$ and $m_\mu/m_\tau\simeq \lambda^2$, where
$\lambda\simeq 0.22$ is the Wolfenstein parameter. Since the
charged lepton mass matrix is of diagonal form, the leptonic mixing
angles stem entirely from the neutrino sector. Enforced by the
symmetries, the vacuum structure yields an exactly maximal atmospheric mixing
angle $\theta_{23}=\pi/4$ and a vanishing reactor angle $\theta_{13}=0$.
In addition, all $\mathcal{CP}$ violation phases vanish due to the
symmetries. The model provides an order of magnitude understanding of the solar
mixing angle $\theta_{12}$, which is predicted to be large but not necessarily close to
maximal. Specifically, without tuning of parameters ({\it i.e.}, by choosing
universal values for all real Yukawa couplings of order unity) the model yields
the solar mixing angle $\theta_{12}={\rm arctan}\:1/\sqrt{2}\simeq 35^\circ$.
The neutrinos exhibit a normal mass hierarchy through single right-handed neutrino
dominance which is realized by
the propagation of right-handed neutrinos in latticized $\mathcal{S}^1/Z_2$
orbifold extra dimensions. Here, the solar mass squared difference
$\Delta m_\odot^2$ is
suppressed against the atmospheric mass squared difference
$\Delta m_{\rm atm}^2$ by the discretized analogue of the volume factor, known
from the classical theory. For a latticization of the orbifold with
$10^1-10^2$ lattice sites the model yields without tuning of parameters the
MSW LMA solution (LMA-I) of the solar neutrino problem within the 90\%
confidence level region.

\section*{Acknowledgements}
I would like to thank A. Falkowski, M. Lindner and T. Ohlsson for useful
comments and discussions. Furthermore, I would like to thank the mathematical
physics department of the Royal Institute of Technology (KTH), Stockholm
(Sweden), for the warm hospitality, where part of this work was done.
This work was supported by the ``Sonderforschungsbereich 375 f\"ur
Astro-Teilchenphysik der Deutschen Forschungsgemeinschaft''.

\appendix
\section{The dihedral group $\mathscr{D}_4$}\label{app:dihedral}
\begin{figure}
\begin{center}
\includegraphics*[bb = 223 530 362 670]{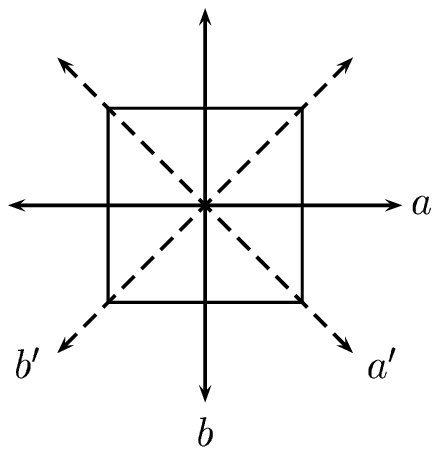}
\end{center}
\vspace{-3mm}
\caption{\small{Horizontal plane with the system of two-fold axes for 
 $\mathscr{D}_4$.}}
\label{fig:dihedral}
\end{figure}

The dihedral groups $\mathscr{D}_n$, where $n=2,3,\ldots $, are the
point-symmetry groups with an $n$-fold axis\footnote{This is also called the
{\it principal} axis.} and a system of 2-fold axes at right
angle to it. The group $\mathscr{D}_n$ is therefore the symmetry group of a
regular $n$-gon. These groups contain $2n$ elements and for $n>2$ they are
non-Abelian ($\mathscr{D}_2$ is isomorphic with the Klein group $Z_2\times Z_2$).
In Fig. \ref{fig:dihedral} the horizontal plane is shown for the case $n=4$,
where $a,a',b,$ and $b'$ denote the four two-fold axes and the 4-fold axis is
perpendicular to the paper.
We denote by $C_n$ the operation of rotation through
$2\pi/n$ about the principal axis. The $k$-fold application of this
transformation will be written as $C_n^k$ and the identity transformation
$C^n_n$ as $E$. The rotations through $\pi$ about the
axes $a,a',b,$ and $b'$ will be referred to as $C_a,C_{a'},C_b,$ and $C_{b'}$,
respectively. Then, the dihedral group ${\mathscr{D}}_4$ has eight elements in
following five classes:
\begin{equation}\label{eq:classes}
 E;\quad C_4,C_4^3;\quad C_4^2;\quad C_a,C_b;\quad C_{a'},C_{b'}.
\end{equation}
Adopting the notation of Ref.~\cite{ham89} we will refer to the five classes
which are associated with the sequence
in Eq.~(\ref{eq:classes}) as $E,C_4(2),C_4^2,C_2(2),$ and $C_{2'}(2)$.
Calling the 4 irreducible singlet representations
$\mathbf{1_A},\mathbf{1_B},\mathbf{1_C},$ and $\mathbf{1_D}$
respectively, the decomposition of the product of the two-dimensional irrep
$\mathbf{2}$ reads
\begin{equation}\label{eq:2x2}
 \mathbf{2}\times\mathbf{2}=
 \mathbf{1_A}+\mathbf{1_B}+\mathbf{1_C}+\mathbf{1_D}.
\end{equation}
The character table for the group ${\mathscr{D}}_4$ is given in table
\ref{tab:characters}.
Denoting $\mathbf{2}$ as $(a,b)$ we have for the singlet representations
\begin{subequations}\label{eq:singlets}
\begin{eqnarray}
 \mathbf{1_A}&=&a_1a_2+b_1b_2,\\
 \mathbf{1_B}&=&a_1b_2-a_2b_1,\\
 \mathbf{1_C}&=&a_1a_2-b_1b_2,\\
 \mathbf{1_D}&=&a_1b_2+a_2b_1.
\end{eqnarray}
\end{subequations}
From the character table of $\mathscr{D}_4$ one determines the
decomposition of the product of any two representations as shown in table
\ref{tab:multiplication}. 
\begin{table}
\begin{center}
\begin{tabular}{|c|c|c|c|c|c|}
\hline
$\mathscr{D}_4$    & $E$ & $C_4^2$ & $C_4(2)$ & $C_2(2)$ & $C_{2'}(2)$ \\ \hline\hline
$\mathbf{1_A}$    & 1 & ~1 & ~1 & ~1 & ~1 \\
$\mathbf{1_B}$  & 1 & ~1 & ~1 & -1 & -1 \\
$\mathbf{1_C}$    & 1 & ~1 & -1 & ~1 & -1 \\
$\mathbf{1_D}$    & 1 & ~1 & -1 & -1 & ~1 \\
$\mathbf{2}$ & 2 & -2 & ~0 & ~0 & ~0 \\
\hline
\end{tabular}
\end{center}
\caption{Character table for the group $\mathscr{D}_4$.}
\label{tab:characters}
\end{table}
\begin{table}
\begin{center}
\begin{tabular}{|c|c|c|c|c|c|}
\hline
$\mathscr{D}_4$   & $\mathbf{1_A}$ & $\mathbf{1_B}$ & $\mathbf{1_C}$ &
 $\mathbf{1_D}$ & $\mathbf{2}$ \\ \hline\hline
$\mathbf{1_A}$  & $\mathbf{1_A}$ & $\mathbf{1_B}$ & $\mathbf{1_C}$ &
 $\mathbf{1_D}$ & $\mathbf{2}$ \\
$\mathbf{1_B}$  &  & $\mathbf{1_A}$ & $\mathbf{1_D}$ & $\mathbf{1_C}$ &
 $\mathbf{2}$ \\
$\mathbf{1_C}$  &  &  & $\mathbf{1_A}$ & $\mathbf{1_B}$ & $\mathbf{2}$ \\
$\mathbf{1_D}$  &  &  &  & $\mathbf{1_A}$ & $\mathbf{2}$ \\
\hline
\end{tabular}
\end{center}
\caption{Multiplication table for the group $\mathscr{D}_4$.}
\label{tab:multiplication}
\end{table}
In the vector representation of ${\mathscr{D}}_4$, the $2\times 2$
representation matrices corresponding to the different classes can be written as
\begin{eqnarray}
E &\quad:\quad& D(E)=
\left(
\begin{matrix}
1 & 0\\
0 & 1
\end{matrix}
\right),\nonumber\\
C_4(2) &\quad:\quad& D(C_4)=
\left(
\begin{matrix}
0 & 1\\
-1 & 0
\end{matrix}
\right),\quad
D(C_4^3)=
\left(
\begin{matrix}
0 & -1\\
1 & 0
\end{matrix}
\right),
\nonumber\\
C_4^2&\quad:\quad&
D(C_4^2)=
\left(
\begin{matrix}
-1 & 0\\
0 & -1
\end{matrix}
\right),\nonumber\\
C_2(2) &\quad:\quad& D(C_a)=
\left(
\begin{matrix}
1 & 0\\
0 & -1
\end{matrix}
\right),\quad
D(C_b)=
\left(
\begin{matrix}
-1 & 0\\
0 & 1
\end{matrix}
\right),
\nonumber\\
C_{2'}(2) &\quad:\quad& D(C_{a'})=
\left(
\begin{matrix}
0 & -1\\
-1 & 0
\end{matrix}
\right),\quad
D(C_{b'})=
\left(
\begin{matrix}
0 & 1\\
1 & 0
\end{matrix}
\right).
\end{eqnarray}
Before concluding this section, let us construct the dihedral groups from
semi-direct products. For this purpose, let $N\simeq Z_n$ for any
$n\in \mathbbm{N}$, let $H\simeq Z_2$,
and let the map $\varphi:H\rightarrow\text{Aut}(N)$ send $h\in H$ to the
automorphism in $\text{Aut}(N)$ sending each element of $N$ to its inverse. Then, the external
semi-direct product $N\rtimes_\varphi H$ of $N$ by $H$ with respect to $\varphi$
is the dihedral group $\mathscr{D}_n$. One can also define the
{\it infinite dihedral group} $\mathscr{D}_\infty = N\rtimes_\varphi H$, where $N\simeq Z$ and the group 
$H$ and the map $\varphi$ are as above.

\section{Minimization of the tree-level potential}\label{app:minimization}
We will rewrite the potential $V_A(\Phi,\Omega)$ in Eqs.~(\ref{eq:V_AV_B}) in terms of the
parameterization in Eqs.~(\ref{eq:parameterization}) as follows
\begin{eqnarray}\label{eq:V_Arewritten}
 V_A
&=&
d_1v_1^4c_\alpha^2s_\alpha^2+d_2v_2^4c_\beta^2s_\beta^2+d_3v_1^2v_2^2
(c_\alpha^2-s_\alpha^2)(c_\beta^2-s_\beta^2),
\end{eqnarray}
where $s_\alpha\equiv{\rm sin}(\alpha)$ and $c_\alpha\equiv{\rm cos}(\alpha)$
(correspondingly for $\beta$). Hence, it is
\begin{eqnarray*}
 \frac{\partial V_A}{\partial \alpha} & = &
2d_1v_1^4(c_\alpha^3s_\alpha-c_\alpha s_\alpha^3)-4d_3v_1^2v_2^2
c_\alpha s_\alpha(c_\beta^2-s_\beta^2),\\
&&\nonumber\\
\frac{\partial V_A}{\partial \beta} & = &
2d_2v_2^4(c_\beta^3s_\beta-c_\beta s_\beta^3)-4d_3v_1^2v_2^2
c_\beta s_\beta(c_\alpha^2-s_\alpha^2).
\end{eqnarray*}
As a result, at $(\alpha,\beta)=(\frac{\pi}{4},\frac{\pi}{4})$ it is
$(\partial_\alpha,\partial_\beta)V_A=(0,0)$,
{\it i.e.}, $(\alpha,\beta)=(\frac{\pi}{4},\frac{\pi}{4})$ is an extremum of the potential $V_A$. For the second
derivatives it follows
\begin{eqnarray*}
 \frac{\partial^2 V_A}{\partial \alpha ^2}&=&
 2d_1v_1^4(s_\alpha^4+c_\alpha^4-6c_\alpha^2s_\alpha^2)
 -4d_3v_1^2v_2^2(c_\alpha^2-s_\alpha^2)(c_\beta^2-s_\beta^2),\\
 &&\nonumber\\
 \frac{\partial^2 V_A}{\partial \beta ^2}&=&
 2d_2v_2^4(s_\beta^4+c_\beta^4-6c_\beta^2s_\beta^2)
 -4d_3v_1^2v_2^2(c_\beta^2-s_\beta^2)(c_\alpha^2-s_\alpha^2),\\
 && \nonumber\\
 \frac{\partial^2 V_A}{\partial \alpha\:\partial \beta}&=&
 16d_3v_1^2v_2^2c_\alpha s_\alpha c_\beta s_\beta.
\end{eqnarray*}
At $(\alpha,\beta)=(\frac{\pi}{4},\frac{\pi}{4})$ we therefore obtain for the matrix
of second derivatives of the potential
\begin{equation}\label{eq:diff^2alphabeta}
 \left(\frac{\partial^2 V_A}{\partial\alpha\:\partial\beta}\right)
=
\left(
\begin{matrix}
 -2d_1v_1^4 & 4 d_3v_1^2v_2^2\\
 4d_3v_1^2v_2^2 & -2d_2v_2^4
\end{matrix}\right),
\end{equation}
where,  due to the choice $d_1,d_2<0$, the diagonal elements are positive.
Hence, if the parameters obey the condition
\begin{equation}\label{eq:range1}
 d_1d_2>4d_3^2,
\end{equation}
the matrix in Eq.~(\ref{eq:diff^2alphabeta}) is positive definite, {\it i.e.},
the modes which oscillate in the $(\alpha,\beta)$-subspace
around $(\alpha,\beta)=(\frac{\pi}{4},\frac{\pi}{4})$ have positive masses and
$(\alpha,\beta)=(\frac{\pi}{4},\frac{\pi}{4})$ is a minimum of the scalar potential $V_A$.
Let us now rewrite the part of the multi-scalar potential $V_B(\Phi,\Omega)$ in
Eqs.~(\ref{eq:V_AV_B}) using the parameterization in Eqs.~(\ref{eq:parameterization})
as follows
\begin{eqnarray}\label{eq:V_B}
V_B
&=& 2d_4v_1^4c_\alpha^2s_\alpha^2{\rm cos}(2\varphi)+
2d_5v_2^4c_\beta^2s_\beta^2\:{\rm cos}(2\psi)\nonumber\\
&&+4d_6v_1^2v_2^2c_\alpha s_\alpha c_\beta s_\beta\:{\rm cos}(\varphi)
\:{\rm cos}(\psi)\nonumber\\
&&-4d_7v_1^2v_2^2c_\alpha s_\alpha c_\beta s_\beta\:{\rm sin}(\varphi)\:
{\rm sin}(\psi),
\end{eqnarray}
where we have used the notation of Eq.~(\ref{eq:V_Arewritten}). Hence, one
concludes
\begin{eqnarray*}
 \frac{\partial V_B}{\partial \alpha}&=&
  4d_4v_1^4(c_\alpha^3s_\alpha-c_\alpha s_\alpha^3)\:{\rm
  cos}(2\varphi)
  +4d_6v_1^2v_2^2(c_\alpha^2-s_\alpha^2)c_\beta s_\beta\:{\rm cos}(\varphi)
  \:{\rm cos}(\psi)\nonumber\\
  &&-4d_7v_1^2v_2^2(c_\alpha^2-s_\alpha^2)c_\beta s_\beta\:
  {\rm sin}(\varphi)\:{\rm sin}(\psi),\\
 \frac{\partial V_B}{\partial \beta}&=&
  4d_5v_2^4(c_\beta^3s_\beta-c_\beta s_\beta^3)\:{\rm
  cos}(2\psi)
  +4d_6v_1^2v_2^2(c_\beta^2-s_\beta^2)c_\alpha s_\alpha\:{\rm cos}(\varphi)
  \:{\rm cos}(\psi)\nonumber \\
  &&-4d_7v_1^2v_2^2(c_\beta^2-s_\beta^2)c_\alpha s_\alpha\:
  {\rm sin}(\varphi)\:{\rm sin}(\psi),
\end{eqnarray*}
and
\begin{eqnarray*}
 \frac{\partial V_B}{\partial \varphi}&=&
  -4d_4v_1^4c_\alpha^2s_\alpha^2{\rm sin}(2\varphi)
  -4d_6v_1^2v_2^2c_\alpha s_\alpha  c_\beta s_\beta\:
  {\rm sin} (\varphi)\:{\rm cos}(\psi)\nonumber\\
  &&-4d_7v_1^2v_2^2c_\alpha s_\alpha c_\beta s_\beta\:{\rm cos}(\varphi)
  \:{\rm sin}(\psi),\\
  \frac{\partial V_B}{\partial \psi}&=&
  -4d_5v_2^4c_\beta^2s_\beta^2{\rm sin}(2\psi)
  -4d_6v_1^2v_2^2c_\alpha s_\alpha  c_\beta s_\beta\:
  {\rm cos} (\varphi)\:{\rm sin}(\psi)\nonumber\\
  &&-4d_7v_1^2v_2^2c_\alpha s_\alpha c_\beta s_\beta\:{\rm sin}(\varphi)
  \:{\rm cos}(\psi).
\end{eqnarray*}
As a result, at the points $(\alpha,\beta)=(\frac{\pi}{4},\frac{\pi}{4})$, where
$\varphi,\psi\in\{0,\pi\}$, it is
\begin{equation*}
(\partial_\alpha,\partial_\beta,\partial_\varphi,\partial_\psi)V_B=(0,0,0,0),
\end{equation*}
{\it i.e.}, these points are extrema of $V_B$. Furthermore, one finds at
$(\alpha,\beta)=(\frac{\pi}{4},\frac{\pi}{4})$ vanishing mixed second derivatives
\begin{equation*}
 \frac{\partial^2 V_B}{\partial\varphi\partial\alpha}=
 \frac{\partial^2 V_B}{\partial\psi\partial\alpha}=
 \frac{\partial^2 V_B}{\partial\varphi\partial\beta}=
 \frac{\partial^2 V_B}{\partial\psi\partial\beta}=0,
\end{equation*}
implying that the matrix of the second derivatives of $V_B$ with respect to
the parameters $\alpha,\beta,\varphi,\psi$ breaks up into a block-diagonal form
with submatrices which respectively correspond to the subspaces $(\alpha,\beta)$
and $(\varphi,\psi)$. The second derivatives of $V_B$ with respect to $\alpha$
and $\beta$ are
\begin{eqnarray*}
 \frac{\partial^2 V_B}{\partial \alpha^2}&=&
 4d_4v_1^4(s_\alpha^4+c_\alpha^4-6c_\alpha^2s_\alpha^2)\:
 {\rm cos}(2\varphi)
 -16d_6v_1^2v_2^2c_\alpha s_\alpha c_\beta s_\beta\:{\rm cos}(\varphi)\:
 {\rm cos}(\psi)\nonumber\\
 &&+16d_7v_1^2v_2^2c_\alpha s_\alpha c_\beta s_\beta\:{\rm sin}(\varphi)\:
 {\rm sin}(\psi),\\
 \frac{\partial^2 V_B}{\partial \beta^2}&=&
 4d_5v_2^4(s_\beta^4+c_\beta^4-6c_\beta^2s_\beta^2)\:
 {\rm cos}(2\psi)
 -16d_6v_1^2v_2^2c_\alpha s_\alpha c_\beta s_\beta\:{\rm cos}(\varphi)\:
 {\rm cos}(\psi)\nonumber\\
 &&+16d_7v_1^2v_2^2c_\alpha s_\alpha c_\beta s_\beta\:{\rm sin}(\varphi)\:
 {\rm sin}(\psi),\\
 \frac{\partial^2 V_B}{\partial \alpha \partial \beta}&=&
 4d_6v_1^2v_2^2(c_\alpha^2-s_\alpha^2)(c_\beta^2-s_\beta^2)\:
 {\rm cos}(\varphi)\:{\rm cos}(\psi)\nonumber\\
 &&-4d_7v_1^2v_2^2(c_\alpha^2-s_\alpha^2)(c_\beta^2-s_\beta^2)\:{\rm sin}
 (\varphi)\:{\rm sin}(\psi).
\end{eqnarray*}
Therefore, at the points $(\alpha,\beta)=(\frac{\pi}{4},\frac{\pi}{4})$,
where $\varphi,\psi\in\{0,\pi\}$, the matrix of the second order derivatives is
\begin{equation}
 \left(\frac{\partial^2 V_B}{\partial\alpha\:\partial\beta}\right)
=
4\left(
\begin{matrix}
  -d_4v_1^4-\sigma d_6v_1^2v_2^2 & 0\\
  0& -d_5v_2^4-\sigma d_6v_1^2v_2^2
\end{matrix}\right),
\end{equation}
where $\sigma\equiv {\rm cos}(\varphi)\:{\rm cos(\psi)}=\pm 1$ can take either sign
for $\varphi,\psi\in\{0,\pi\}$. However, from Eq.~(\ref{eq:V_B}) it is seen that
the product $d_6\sigma$ must be negative in the lowest energy state, {\it i.e.},
the sign of $d_6$ determines whether $\varphi=\psi+k\cdot2\pi$ or
$\varphi=\psi+k\cdot\pi$ for
some integer $k$. The matrix of second order derivatives can therefore be
rewritten as
\begin{equation}
 \left(\frac{\partial^2 V_B}{\partial\alpha\:\partial\beta}\right)
=
4\left(
\begin{matrix}
  -d_4v_1^4+|d_6|v_1^2v_2^2 & 0\\
  0& -d_5v_2^4+|d_6|v_1^2v_2^2
\end{matrix}\right),
\end{equation}
where $d_4,d_5<0$, {\it i.e.}, the matrix is positive definite.  
 The second derivatives of $V_B$ with respect to $\varphi$ and $\psi$ are
\begin{eqnarray*}
 \frac{\partial^2 V_B}{\partial\varphi^2}&=&
 -8d_4v_1^4c_\alpha^2s_\alpha^2\:{\rm cos}(2\varphi)\nonumber\\
 &&-4d_6v_1^2v_2^2c_\alpha s_\alpha c_\beta s_\beta \:{\rm cos}(\varphi)\:
 {\rm cos}(\psi)
 +4d_7v_1^2v_2^2c_\alpha s_\alpha c_\beta s_\beta \:{\rm sin}(\varphi)\:
 {\rm sin}(\psi),\\
 \frac{\partial^2 V_B}{\partial\psi^2}&=&
 -8d_5v_2^4c_\beta^2s_\beta^2\:{\rm cos}(2\psi)\nonumber\\
 &&-4d_6v_1^2v_2^2c_\alpha s_\alpha c_\beta s_\beta \:{\rm cos}(\varphi)\:
 {\rm cos}(\psi)
 +4d_7v_1^2v_2^2c_\alpha s_\alpha c_\beta s_\beta \:{\rm sin}(\varphi)\:
 {\rm sin}(\psi),\\
 \frac{\partial^2 V_B}{\partial \psi \partial \varphi}&=&
 4d_6v_1^2v_2^2c_\alpha s_\alpha c_\beta s_\beta \:{\rm sin}(\varphi)\:
 {\rm sin}(\psi)
 -4d_7v_1^2v_2^2c_\alpha s_\alpha c_\beta s_\beta\: {\rm cos}(\varphi)\:
 {\rm cos}(\psi).
\end{eqnarray*}
At the points $(\alpha,\beta)=(\frac{\pi}{4},\frac{\pi}{4})$, where
$\varphi,\psi\in\{0,\pi\}$, the matrix of the second derivatives of $V_B$ with
respect to $\varphi$ and $\psi$ reads
\begin{equation}\label{eq:diff^2phipsi}
 \left(\frac{\partial^2 V_B}{\partial\varphi\partial\psi}\right)=
 \left(
 \begin{matrix}
  -2d_4v_1^4+|d_6|v_1^2v_2^2 & \pm d_7v_1^2v_2^2\\
  \pm d_7v_1^2v_2^2&-2d_5v_2^2+|\sigma| d_6v_1^2v_2^2
 \end{matrix}
 \right),¶
\end{equation}
where $d_4,d_5<0$, {\it i.e.}, the diagonal elements are
positive. In Eq.~(\ref{eq:diff^2phipsi}) we have already used
that the potential is minimized when $d_6\sigma$ is negative.

Taking everything into account, if the
coefficients in the multi-scalar potential satisfy besides
Eq.~(\ref{eq:range1}) also the condition
\begin{equation}\label{eq:range2}
 (-2d_4v_1^4+|d_6|v_1^2v_2^2)
 (-2d_5v_2^4+|d_6|v_1^2v_2^2)>d_7^2v_1^4v_2^4,
\end{equation}
then the matrix in Eq.~(\ref{eq:diff^2phipsi}) is positive definite,
{\it i.e.}, all modes oscillating in the $(\alpha,\beta,\varphi,\psi)$-subspace
around the points $(\alpha,\beta)=(\frac{\pi}{4},\frac{\pi}{4})$,
where $\varphi,\psi\in\{0,\pi\}$, have positive masses and hence these points are
indeed local minima of both the potentials $V_A$ and $V_B$, {\it i.e.}, they
locally minimize the term $V_\Delta(\Phi,\Omega)$ which breaks the accidental
$U(1)^4_{\rm acc}$-symmetry.

\end{document}